\documentclass[a4paper,fleqn,usenatbib,useAMS]{mnras}
\usepackage{newtxtext,newtxmath}
\usepackage[T1]{fontenc}
\usepackage{ae,aecompl}
\usepackage{graphicx}
\usepackage{natbib}
\usepackage{amssymb}
\usepackage{amsmath}
\usepackage{pdflscape}
\usepackage{xcolor}

\def\lesssim{\lower.5ex\hbox{$\; \buildrel < \over \sim \;$}}
\def\gtrsim{\lower.5ex\hbox{$\; \buildrel > \over \sim \;$}}

\title[Intrinsic alignments of high-$z$ elongated low-mass galaxies]{Can intrinsic alignments of elongated low-mass galaxies be used to map the cosmic web at high redshift?}

\author[V. Pandya et al.] {Viraj Pandya$^1$\thanks{email: viraj.pandya@ucsc.edu}, Joel Primack$^2$, Peter Behroozi$^3$, Avishai Dekel$^4$, Haowen Zhang$^3$, 
\newauthor Elliot Eckholm$^1$, Sandra M. Faber$^1$, Henry C. Ferguson$^5$, Mauro Giavalisco$^{6}$,
\newauthor Yicheng Guo$^7$, Nimish Hathi$^5$, Dritan Kodra$^8$, Anton M. Koekemoer$^5$,
\newauthor David C. Koo$^1$, Jeffrey Newman$^8$ and Arjen van der Wel$^{9,10}$\\
$^1$UCO/Lick Observatory, Department of Astronomy and Astrophysics, University of California, Santa Cruz, CA 95064, USA\\
$^2$Department of Physics, University of California at Santa Cruz, Santa Cruz, CA 95064, USA\\
$^3$Department of Astronomy and Steward Observatory, University of Arizona, Tucson, AZ 85721, USA\\
$^4$Center for Astrophysics and Planetary Science, Racah Institute of Physics, The Hebrew University, Jerusalem 91904, Israel\\
$^5$Space Telescope Science Institute, 3700 San Martin Drive, Baltimore, MD 21218, USA\\
$^6$Department of Astronomy, University of Massachusetts Amherst, 710 North Pleasant Street, Amherst, MA 01003-9305, USA\\
$^7$Department of Physics and Astronomy, University of Missouri, Columbia, MO 65211, USA\\
$^8$Department of Physics and Astronomy and PITT PACC, University of Pittsburgh, Pittsburgh, PA 15260, USA\\
$^{9}$Sterrenkundig Observatorium, Universiteit Gent, Krijgslaan 281 S9, B-9000 Gent, Belgium\\
$^{10}$Max-Planck Institut f\"{u}r Astronomie, K\"{o}nigstuhl 17, D- 69117, Heidelberg, Germany\\
}

\date{Accepted ???. Received ??? in original form ???}

\begin{document}
\label{firstpage}
\pagerange{\pageref{firstpage}--\pageref{lastpage}}
\maketitle

\begin{abstract}
Hubble Space Telescope observations show that low-mass ($M_*=10^9-10^{10}M_{\odot}$) galaxies at high redshift ($z=1.0-2.5$) tend to be elongated (prolate) rather than disky (oblate) or spheroidal. This is explained in zoom-in cosmological hydrodynamical simulations by the fact that these galaxies are forming in cosmic web filaments where accretion happens preferentially along the direction of elongation. We ask whether the elongated morphology of these galaxies allows them to be used as effective tracers of cosmic web filaments at high redshift via their intrinsic alignments. Using mock lightcones and spectroscopically-confirmed galaxy pairs from the CANDELS survey, we test two types of alignments: (1) between the galaxy major axis and the direction to nearby galaxies of any mass, and (2) between the major axes of nearby pairs of low-mass, likely prolate, galaxies. The mock lightcones predict strong signals in 3D real space, 3D redshift space, and 2D projected redshift space for both types of alignments (assuming prolate galaxy orientations are the same as those of their host prolate halos), but we do not detect significant alignment signals in CANDELS observations. However, we show that spectroscopic redshifts have been obtained for only a small fraction of highly elongated galaxies, and accounting for spectroscopic incompleteness and redshift errors significantly degrades the 2D mock signal. This may partly explain the alignment discrepancy and highlights one of several avenues for future work.
\end{abstract}


\section{Introduction}\label{sec:intro}
There is observational evidence for the idea that low-mass galaxies at high redshift start out with intrinsically elongated (prolate) shapes rather than being disky (oblate) or spheroidal \citep[e.g.,][]{cowie95,ravindranath06,yuma12,law12}. \citet{vanderwel14} used the distribution of projected axis ratios measured with the Hubble Space Telescope (HST) for galaxies at $z\sim2$ in the Cosmic Assembly Near-infared Deep Extragalactic Legacy Survey \citep[CANDELS;][]{candels1,candels2} to infer that the intrinsic shapes of the majority of low-mass, high-redshift galaxies are prolate. More recently, \citet{zhang18} extended the analysis of \newpage\noindent\citet{vanderwel14} by simultaneously comparing the distributions of both projected axis ratios and semi-major axis lengths of CANDELS galaxies to the expectation from projecting ellipsoids with a range of intrinsic axis ratios and sizes (oblate, prolate and spheroidal). \citet{zhang18} found that $>50\%$ of observed CANDELS galaxies with stellar masses $10^9-10^{10}M_{\odot}$ at $1.0<z<2.5$ are likely to be intrinsically prolate (with typical projected axis ratios of $b/a\sim0.3$); at lower redshifts, the prolate fraction drops considerably due to the emergence of oblate and spheroidal systems. 

On the theoretical side, \citet{ceverino15} used hydrodynamical cosmological ``zoom-in" simulations to investigate the intrinsic shapes of low-mass galaxies. They found that galaxies in their simulations were intrinsically prolate at low masses and early times. The host halos of these prolate galaxies were also prolate, implying that they were still forming in cosmic web filaments and hence their accretion of matter was preferentially along the direction of the filament. After the galaxies underwent sufficient central star formation, they became stellar mass dominated in their centers rather than dark matter (DM) dominated, and their stellar mass distribution transitioned from prolate to more typical spheroidal/oblate shapes. Building on \citet{ceverino15}, \citet{tomassetti16} studied whether the direction of elongation of the stellar mass distribution was aligned with the direction of elongation of the dark matter halo itself on scales of $2R_{\rm vir}$. \citet{tomassetti16} found strong intrinsic galaxy--halo alignment for nearly all of their simulated galaxies while in the early dark matter-dominated prolate phase: less than 20\% of their 34 ``zoom-in" simulated galaxies had $\cos\theta<0.9$ when comparing the directions of the longest axis of the dark matter and stellar mass distributions (see their Figure 10).

On even larger scales, intrinsic halo--halo and galaxy--galaxy principal axis alignments are predicted to be ubiquitous within the $\Lambda$CDM cosmology, and these intrinsic alignments themselves reflect an even stronger underlying alignment with the filamentary structure of the cosmic web \citep[e.g., see the recent reviews by][and references therein]{kiessling15,kirk15,joachimi15,troxel15}.\footnote{There are also predictions and observational constraints for small-scale alignments, such as for central--satellite systems \citep[e.g.,][]{brainerd05,yang06,faltenbacher07}. However, in this paper we only focus on larger scale intrinsic alignments, on scales of several comoving Mpc.} Using N-body simulations of $\Lambda$CDM (without baryons), several studies have converged on the prediction that both the shapes and kinematics of halos exhibit alignments with cosmic web filaments, and that shape-based alignments are more robust than kinematic alignments \citep[e.g.,][]{allgood06,aragoncalvo07,libeskind13,foreroromero14}. Two key predictions that seem to be robust to different N-body simulation methods, halo finders, and cosmic web feature identifiers are that: (1) the longest shape axis of a halo tends to be aligned in the direction of a filament regardless of the halo mass, and (2) low-mass halos tend to spin parallel to the filament whereas high mass halos spin perpendicular to the filamentary axis \citep[e.g.,][and references therein]{libeskind13,kiessling15,libeskind18}. As discussed in sections 5 and 6 of \citet{kiessling15}, studies of intrinsic alignments of galaxies in hydrodynamical simulations and semi-analytic models are still sparse owing to the large computational expenses and the uncertainties of subgrid physics \citep[but see, e.g.,][]{tenneti15,chisari15,chen15,codis18}. 

In contrast, observational constraints on intrinsic alignments of galaxies are more sparse and uncertain \citep[see the recent observational review of intrinsic alignments by][]{kirk15}. There have been several independent detections of intrinsic alignments for luminous red galaxies \citep[LRGs; e.g., see][]{mandelbaum06,hirata07,okumura09,joachimi11,singh15}, extending the initially striking discovery by \citet{binggeli82} that the central LRGs of nearby clusters tend to point towards each other. However, to date, the search for intrinsic alignments among blue/disky galaxies has yielded null or insignificant results \citep{kirk15}. Most studies have focused on $z\sim0.1$ owing to the lack of large sample sizes and sufficiently high-quality measurements of shape, redshift, etc. at higher redshifts. A few notable exceptions are \citet{joachimi11}, who detected the expected signal for LRGs out to $z\sim0.7$, and \citet{mandelbaum11} and \citet{tonegawa18}, who extended the null result for blue/disky galaxies out to $z\sim0.6$ and $z\sim1.4$, respectively. Additional high-redshift observational constraints (with an emphasis on weak gravitational lensing) are also now becoming possible with data from upcoming large-area surveys such as the Dark Energy Survey \citep[][who find evidence for decreasing intrinsic alignments among LRGs out to $z\sim1$]{samuroff18}, and the Subaru Hyper Suprime-Cam survey \citep{hikage18}.

Here we test whether elongated, low-mass galaxies at high-redshift can serve as clean tracers of intrinsic alignments for blue galaxies, as might be expected within the $\Lambda$CDM cosmology if these galaxies are indeed actively forming along filaments of the cosmic web \citep{ceverino15,tomassetti16}. Our analysis is enabled by state-of-the-art observational catalogs from the CANDELS survey, and by complementary mock simulation lightcones that mimic the CANDELS survey area and that allow us to predict the expected signal. Our work extends beyond previous studies of intrinsic alignments in three ways: 

\begin{enumerate}
\item we are extending to much higher redshifts ($z=1.0-2.5$) than has typically been observationally probed before 
\item we are extending to lower stellar masses ($\log M_*/M_{\odot}=9-10$) for star-forming galaxies at these higher redshifts
\item unlike previous studies that focus on blue/disky versus LRG/elliptical galaxies, we specifically focus on prolate (elongated) galaxies
\end{enumerate}

This paper is organized as follows. In \autoref{sec:data}, we describe the mock lightcone and observational datasets. In \autoref{sec:analysis} we discuss our analysis procedures for the mock lightcones and observations. We present our results on intrinsic alignments in \autoref{sec:results}. After a discussion in \autoref{sec:discussion}, we conclude with a summary and brief remarks on future observational and theoretical prospects in \autoref{sec:summary}. In Appendix \ref{sec:appcorrfunc}, we compute the angular two-point correlation function for low-mass CANDELS galaxies in four redshift intervals and compare to that of the mock lightcones. We assume the \citet{planck15} cosmology throughout, with $\Omega_{\rm m,0}=0.307$ and $h=0.678$ as in the Bolshoi--Planck simulation \citep{rodriguezpuebla16}. 

\section{Data}\label{sec:data}
Here we describe our mock lightcone and observational datasets. 

\subsection{Mock simulation lightcones}\label{sec:datamocks}
To predict whether an intrinsic alignment signal should exist within the underlying $\Lambda$CDM cosmology, we use mock simulation lightcones from the UniverseMachine subhalo abundance matching model \citep{behroozi18}. These mock lightcones are generated to roughly mimic the cosmological volumes probed by the five CANDELS fields plus ancillary observational data. The underlying sample of halos comes from the Bolshoi-Planck dark matter-only dissipationless simulation \citep{klypin16,rodriguezpuebla16}, which used the cosmological parameters of the \citet{planck15} cosmology. The Bolshoi-Planck cube is 250 Mpc/h on a side with 2048$^3$ dark matter particles and a particle mass resolution of $1.5\times10^8M_{\odot}/h$. As described in detail in \citet{rodriguezpuebla16}, the halo properties were determined using the Rockstar halo finder \citep{behroozi13rockstar} and merger trees were extracted using the \texttt{consistent-trees} code \citep{behroozi13ctrees}. We consider both distinct halos and subhalos (satellites) in our analysis because we do not have central/satellite classifications for our observed high-redshift CANDELS galaxies. Given the resolution of the Bolshoi-Planck simulation, we only consider halos with log $M_{\rm vir}/M_{\odot}>11$ to prevent resolution issues. This halo mass limit should include most, if not all, galaxies with $\log M_*/M_{\odot}\geq9$ \citep{behroozi18}.

Eight mock light cones were generated for each of the five CANDELS fields, yielding forty mock light cones in total. For each mock, one halo in the $z=0$ snapshot is randomly chosen to be the ``home galaxy" of the observer, and then a light cone with a solid angle mimicking one CANDELS field (including additional area to allow testing of edge effects on sample selection) is chosen to point in a random direction on the sky. Two rotation matrices are computed for: (1) transforming halo properties from the original box reference frame to the 3D light cone reference frame, and (2) to further transform from the 3D light cone reference frame to an ``on-sky" projected reference frame.

\begin{figure} 
\begin{center}
\includegraphics[width=\hsize]{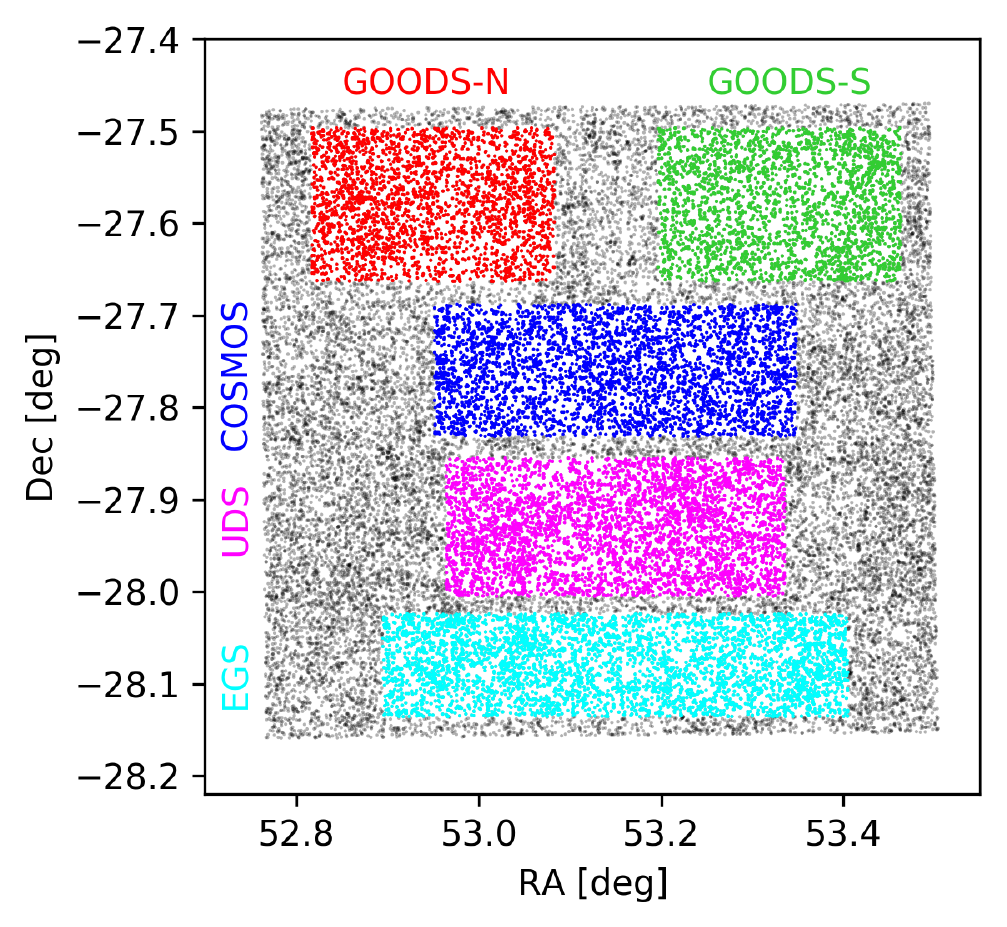}
\end{center}
\caption{The footprint of our overall mock lightcone (black points) with our five individual CANDELS-sized mock subfields overlaid. GOODS-N is in red, GOODS-S in green, COSMOS in blue, UDS in magenta, and EGS in cyan.}
\label{fig:subfields}
\end{figure}

In this work, we opt to use a single mock light cone for the sake of convenience.\footnote{Specifically, we use survey\_GOODS-S\_z0.00-10.00\_x39.00\_y41.00\_0.dat which is available at \url{http://behroozi.users.hpc.arizona.edu/UniverseMachine/EDR/Lightcones/}.} This one mock light cone is $44.5'\times41.3'$ and so has double the on-sky area of all five CANDELS fields combined: $\sim0.51$ deg$^2$ versus $\sim0.26$ deg$^2$. Hence, from our large roughly square mock, we extract five subfields whose rectangular dimensions are equivalent to those of the CANDELS fields: $16'\times10'$ for GOODS-N and GOODS-S, $23.8'\times8.6'$ for COSMOS, $22.3'\times9'$ for UDS, and $30.6'\times6.7'$ for EGS (see the catalog papers referenced below). These mock subfields are illustrated in \autoref{fig:subfields}. 

\subsection{CANDELS observations}
To test whether the expected signal exists, we use existing state-of-the-art observational catalogs from the CANDELS survey for all five CANDELS fields. The backbone for CANDELS is F160W ($H$) band imaging taken with the \textit{Wide Field Camera 3 (WFC3)} aboard the Hubble Space Telescope. The details of source detection, photometric measurements and cataloguing are described in \citet[][COSMOS]{nayyeri17}, \citet[][EGS]{stefanon17}, Barro et al. (in preparation, GOODS-N), \citet[][GOODS-S]{guo13}, and \citet[][UDS]{galametz13}. We impose basic photometric flag cuts on the source catalog for each field, namely: (1) \texttt{PhotFlag==0}, (2) \texttt{CLASS\_STAR<0.8}, and (3) \texttt{WFC3\_F160W\_FLUX>0}. We use stellar mass catalogs based on optical--NIR spectral energy distribution (SED) fitting; the associated systematic uncertainties are described in \citet{santini15} and \citet{mobasher15}. Finally, we only use galaxies that have reliable GALFIT \citep{peng} measurements in the $H$-band \citep{vanderwel12}. 

Our adopted technique for selecting galaxy pairs using 3D comoving pair separations cannot be applied to galaxies with photometric redshifts because the photometric redshift uncertainties are prohibitive. For example, according to \citet{dahlen}, the typical $\Delta z_{\rm phot}\approx0.05(1+z)$ for CANDELS, and this would translate to $\Delta z_{\rm phot}\approx0.1$ for a galaxy at $z=1$. The difference in comoving distance to a galaxy at $z=1$ versus $z=1.1$ is a staggering $\sim240$ cMpc, much larger than our maximum probed pair separations of 10 cMpc. In contrast, spectroscopic redshifts are much more securely determined and should translate to comoving distance uncertainties of at most a few cMpc. Hence, we only consider galaxies that are spectroscopically-confirmed. We use the compilation of reliable spectroscopic redshifts from Kodra et al. (in preparation). We only use spectroscopic redshifts with quality flags of ``good" or ``intermediate" (this means that HST grism redshifts are included). Roughly half of our spectroscopic redshifts are grism redshifts; in \autoref{sec:incompleteness}, we discuss grism redshift uncertainties. 

\section{Analysis}\label{sec:analysis}
Here we describe how we analyzed the datasets, namely how we selected elongated galaxies/halos, defined pairs, and computed alignment angles in the observations and the mocks. 

\subsection{Mock simulation lightcones}
\subsubsection{Selecting prolate halos/galaxies}
Since our mock lightcones are constructed using dark matter-only simulations, we rely on a combination of halo properties and subhalo abundance matching to select pairs in which at least one of the two halos is likely to host an elongated galaxy with low stellar mass. We use the observed stellar masses assigned to halos in the UniverseMachine mock lightcone catalog \citep[these stellar masses account for both intrinsic and observational scatter in the stellar-to-halo-mass relation][]{behroozi18}. First, we select halos in the following four stellar mass--redshift bins, in which \citet{zhang18} found that $>50\%$ of CANDELS galaxies were likely to be intrinsically prolate: 

\begin{enumerate}
\item $z=1.0-1.5$ and $\log M_*=9.0-9.5$
\item $z=1.5-2.0$ and $\log M_*=9.0-9.5$
\item $z=2.0-2.5$ and $\log M_*=9.0-9.5$
\item $z=2.0-2.5$ and $\log M_*=9.5-10.0$
\end{enumerate}

While the CANDELS observations suggest that galaxies in these stellar mass--redshift bins are predominantly prolate, it is worthwhile to further impose a selection criterion on the prolateness probability. \citet{ceverino15} and \citet{tomassetti16} found, using the VELA hydrodynamical zoom-in simulations, that prolate galaxies tend to live in the most prolate halos, which is consistent with \citet{allgood06} who showed that the most prolate halos are low-mass and are still preferentially accreting matter along filaments. Following common practice, we compute the ``triaxiality" parameter for every halo in the four stellar mass--redshift bins above: 
\begin{equation}
T = \frac{1-q^2}{1-s^2}\;,
\end{equation}
where $q=\frac{b}{a}$ and $s=\frac{c}{a}$ are two axis ratios that define the shape of an ellipsoid ($q$ and $s$ are provided in the halo snapshot catalogs). It is common practice to consider halos with $T>2/3$ as prolate, $T<1/3$ as oblate/disky, and $1/3<T<2/3$ as triaxial \citep[see equation 12 in][]{allgood06}. Hence, to obtain our final sample of halos that are most likely to host prolate galaxies, we select only halos in the above stellar mass--redshift bins with $T>2/3$. We emphasize that our conclusions remain the same if we do not apply this triaxiality cut. 

Finally, since our mocks are dark matter-only, we need to make a strong assumption about how halo orientation relates to galaxy orientation in terms of the longest axis. \citet{tomassetti16} found, again using the same VELA hydrodynamical zoom-in simulations, that when halos/galaxies were in the prolate phase, the longest axis of the stellar mass distribution was very well aligned with the longest axis of the dark matter distribution. Indeed, Figure 10 (top-left panel) from \citet{tomassetti16} shows that the cumulative distribution function of the dot product between the longest-axis vectors for the stars and DM is strongly peaked at $\cos\theta\sim1$, with only $<20\%$ of prolate halos/galaxies having $\cos\theta<0.9$. Hence, for simplicity we will make the extreme assumption that our prolate halos would host prolate galaxies and that the longest axis of each would point in the same direction. The Rockstar halo snapshot catalogs provide a ``shape vector" giving the direction of the longest axis of every halo measured within R$_{\rm vir}$ according to the ellipsoid fitting method of \citet{allgood06}. We use this shape vector to determine the direction of elongation of every halo, and hence of the stellar mass distribution assigned to them.

\subsubsection{Defining halo--halo pairs and computing alignment angles}
\begin{figure} 
\begin{center}
\includegraphics[width=0.9\hsize]{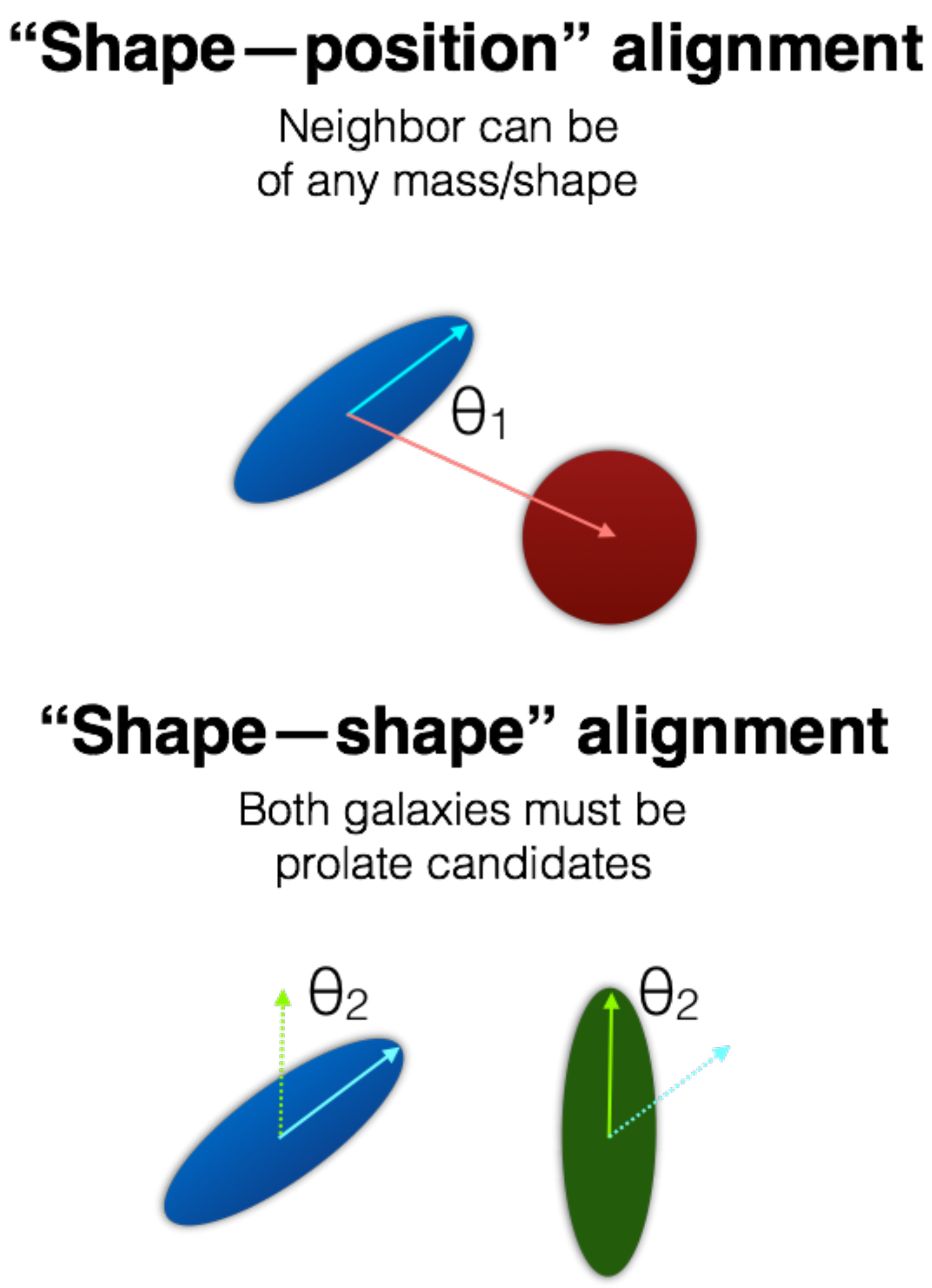}
\end{center}
\caption{An illustration of the two types of alignments we consider in this paper. Top: ``shape--position" alignments are computed between the direction of elongation of the prolate object and the direction to a neighbor (of any mass/shape). Bottom: ``shape--shape" alignments test whether the directions of elongation for two prolate objects themselves point in the same direction (naturally this is only computed for pairs where both objects are considered likely to be prolate). In the 3D mocks, the alignments are computed as dot products between two 3D vectors, whereas in the 2D observations and mocks, the alignments are computed as the difference between two projected position angles.}
\label{fig:illustration}
\end{figure}

With our sample of prolate halos in hand, we now need to define halo--halo pairs. We consider two types of alignments that we illustrate in \autoref{fig:illustration}. For ``shape--position" alignments, we simply want to know whether the longest axis of a DM halo points in the direction to a neighbor of any mass, whereas for ``shape--shape" alignments, we want to know whether the direction of longest axes of two prolate halos themselves point in the same direction (e.g., tracing out a filamentary chain of halos). Hence, for ``shape--shape alignments" we only consider neighboring halos that are also likely to host prolate galaxies (i.e., they surpass the stellar mass--redshift and triaxiality cuts above), whereas for ``shape--position alignments" we do not impose any mass/shape cut for the neighbors (except log $M_{\rm vir}>11$ to prevent resolution issues). 

For ``shape--position" alignments, we compute the vector pointing from the center of each low-mass prolate halo in our sample to each of its neighbors within 10 comoving Mpc (cMpc). This comoving pair separation threshold of 10 cMpc is rather large in proper distance units (10 cMpc corresponds to 5 pMpc at $z=1$ and 3.3 pMpc at $z=2$), but it helps with sample statistics in the observations. It also enables us to show our alignment angles as a function of pair separation for nearest neighbors only and also when including all neighbors within 10 cMpc. In 3D, we can compute the alignment angle by taking the dot product between two vectors: 
\begin{equation}
|\cos\theta| = \frac{\left|v_1 \cdot v_2\right|}{||v_1|| \; ||v_2||} \;,
\end{equation}
where $\theta$ is the plane alignment angle between the two vectors $v_1$ and $v_2$, and the absolute value is taken because we only want to know whether the two vectors are parallel or perpendicular (hence $|\cos\theta|$ will have values between 0 and 1). For ``shape--position" alignments, the relevant vectors for each pair are the shape vector of the low-mass prolate halo and the position difference vector between the two halos. For ``shape--shape" alignments, the relevant vectors are the shape vectors of both low-mass prolate halos. For both kinds of alignments, the 3D comoving pair separation is computed simply as the norm of the position difference vector.

In the mocks, we can measure the expected intrinsic alignments in 3D real space, and then do two degradations: from 3D real space to 3D redshift space, and then from 3D redshift space to 2D projected redshift space. The first degradation will allow us to estimate how redshift space distortions (RSDs) weaken intrinsic ``shape--position" alignments since the position difference vectors will change (shape--shape alignments should be relatively robust to RSDs). The second degradation will allow us to transform our results into the observational plane and ask if any net intrinsic alignments we expect in 3D redshift space will still be observable in 2D projection. The degradation from 3D real space to 3D redshift space simply involves scaling the real position vectors of halos by the ratio of comoving distances to $z_{\rm los}$ and $z_{\rm cosmo}$, where $z_{\rm los}$ is the redshift including peculiar velocities and $z_{\rm cosmo}$ is the redshift due solely to the Hubble flow \citep[both of these redshifts are given in the UniverseMachine lightcone catalogs;][]{behroozi18}. 

For 2D projection, we use the standard transformation equations to go from 3D Cartesian coordinates to spherical polar coordinates, which gives us the radial (line-of-sight) distance and the on-sky (projected) RA and Dec of the halo center and another position along the direction of the halo's longest axis. We quantify the 2D projected halo shape by computing the position angle between the (RA, Dec) of the halo center and the additional on-sky position along the direction of the longest axis. Then, just as for the observations, in 2D projected redshift space, we compute comoving pair separations by converting the (RA, Dec) and redshift of two halo centers to 3D Cartesian coordinates, and taking the norm of the resulting position difference vector. Unlike in 3D, alignment angles are computed as the difference between two on-sky position angles, and we denote this difference $\psi$.\footnote{In 3D, as is common practice, we report $\cos\theta$ instead of $\theta$ because $\theta$ is the polar angle in 3D spherical coordinates. A uniform distribution of points on the surface of a sphere will be uniform in $\cos\theta$ but not $\theta$ because there is less area at the poles of the sphere. Since we will be comparing our distribution of measured alignment angles to the expectation from a uniform distribution, it is hence natural to use $\cos\theta$ in 3D. However, this argument does not apply after 2D projection onto a plane, and hence for the 2D mocks and observations we will report $\psi$, the on-sky difference of two position angles.}

In the end, we have 8277 halos in 3D real space that satisfy our selection criteria, yielding 132214 unique halo--halo pairs for shape--position alignments and 40976 unique halo--halo pairs for shape--shape alignments. The numbers of halos and unique pairs are comparable for the mocks in 3D redshift space and 2D projected redshift space. These numbers of halos and pairs are also roughly on par with the combined photometric and spectroscopic CANDELS dataset (indeed this is by design since the mocks were constructed using subhalo abundance matching; we compare the mock and observed correlation functions in Appendix \ref{sec:appcorrfunc}). However, our spectroscopic pair catalog has a smaller sample size than the mock due to spectroscopic incompleteness; we describe how we degrade the mock further to account for this in \autoref{sec:incompleteness}. 

\subsection{Observations}
In the observations, we are restricted to working with 2D projected positions, axis ratios, and alignment angles, and we cannot correct for RSDs. First, we create our sample of elongated low-mass galaxy candidates by selecting galaxies in the four stellar mass--redshift bins where \citet{zhang18} estimated $>50\%$ prolate fractions. Then, we assign each of these individual galaxies probabilities of being intrinsically prolate/elongated, oblate/disky or spheroidal by essentially interpolating the results of \citet{zhang18}. Specifically, we consider the fractional contribution of prolate, oblate and spheroidal ellipsoids for an individual galaxy's combination of projected $b/a$ and semi-major axis length. The ellipsoidal fractions come from modeling by \citet{zhang18} of the $b/a$ versus size diagram using the entire population of galaxies in the stellar mass--redshift bin that the individual galaxy belongs to. We discard all galaxies for which the prolateness probability is $<50\%$; this discards objects with, e.g., high $b/a$ and small size that are likely spheroidal rather than prolate. We emphasize that our conclusions remain the same if we use the full sample of galaxies in the four redshift--stellar mass bins without any cut on prolateness probability. 

For each of our prolate galaxies, we compute 3D comoving pair separations to other galaxies in the catalog by converting the on-sky (RA, Dec) and redshifts to Cartesian coordinates, and then taking the norm of the resulting 3D position difference vector. To maximize sample size and enable us to study alignment angles as a function of pair separation, we consider all neighbors within 10 cMpc (just like for the mocks). For the shape--position alignments sample, we consider all neighbors with log $M_*/M_{\odot}>9$ without any further restriction on their stellar masses and shapes. For the shape--shape alignments sample, we require the neighbors to also have $M_*/M_{\odot}=9-10$ and prolateness probability $>50\%$. 

In this paper, we do not attempt to analyze galaxy pairs in individual mass and redshift bins, and in each CANDELS field separately, due to small number statistics. Instead, we group the unique galaxy pairs from different fields and redshift--mass bins together, which gives us a large enough sample to begin exploring the intrinsic alignments of pairs of high-redshift galaxies. We have 1022 low-mass prolate galaxies with at least one neighbor of any mass within 10 cMpc, and 715 of these prolate galaxies have at least one other low-mass prolate galaxy within 10 cMpc. In total, we have 7669 unique shape--position pairs and 1931 shape--shape pairs (considering all neighbors within 10 cMpc). These numbers of galaxies/pairs are much lower than in our fiducial mock due to spectroscopic incompleteness, which we address below in \autoref{sec:incompleteness}. 

For every low-mass prolate galaxy, we record the 2D projected position angle of its direction of elongation as measured by GALFIT in the $H$-band \citep[given in the catalog from][]{vanderwel12}. We also compute the 2D projected position angle from the (RA, Dec) of every low-mass prolate galaxy candidate to the (RA, Dec) of each of its neighbors within 10 cMpc. Then, the ``shape--position" alignment angle is the difference between the GALFIT position angle of the galaxy shape, and the position angle to a neighbor. The ``shape--shape" alignment angle is simply the difference between the GALFIT position angle of the shapes of the two low-mass prolate galaxies. These 2D projected alignment angles $\psi$ are self-consistently defined to lie within $[0,\frac{\pi}{2}]$ with 0 indicating perfect alignment and $\frac{\pi}{2}$ indicating perpendicularity.

\subsection{Accounting for redshift errors and spectroscopic incompleteness}\label{sec:incompleteness}
Our fiducial 2D mock contains many more pairs overall (especially at small separations) compared to our observational sample. This is due to at least two effects: (1) our observed spectroscopic sample represents only a small fraction of the full photometric CANDELS source catalogs, and (2) roughly half of our observed galaxies have grism-based redshifts whose errors can still be on the order of $\sim10$ cMpc, leading to a bias against very small separation pairs. We address these issues by creating a suite of ``spectroscopically-degraded" 2D mocks where we mock grism redshift errors for 50\% of the halos (randomly selected), and then randomly draw the same number of mock pairs as a function of pair separation as in the observations. We repeat the random subsampling step 1000 times to assess the probability that any given random subsample will predict a strong alignment signal (separately for nearest neighbor pairs and pairs including all neighbors out to 10 cMpc, for both alignment scenarios). For the grism redshift errors, we assume a Gaussian with mean zero and standard deviation of $\Delta z_{\rm grism}=0.003(1+z)$, following \citet{momcheva16}.

\begin{figure} 
\begin{center}
\includegraphics[width=\hsize]{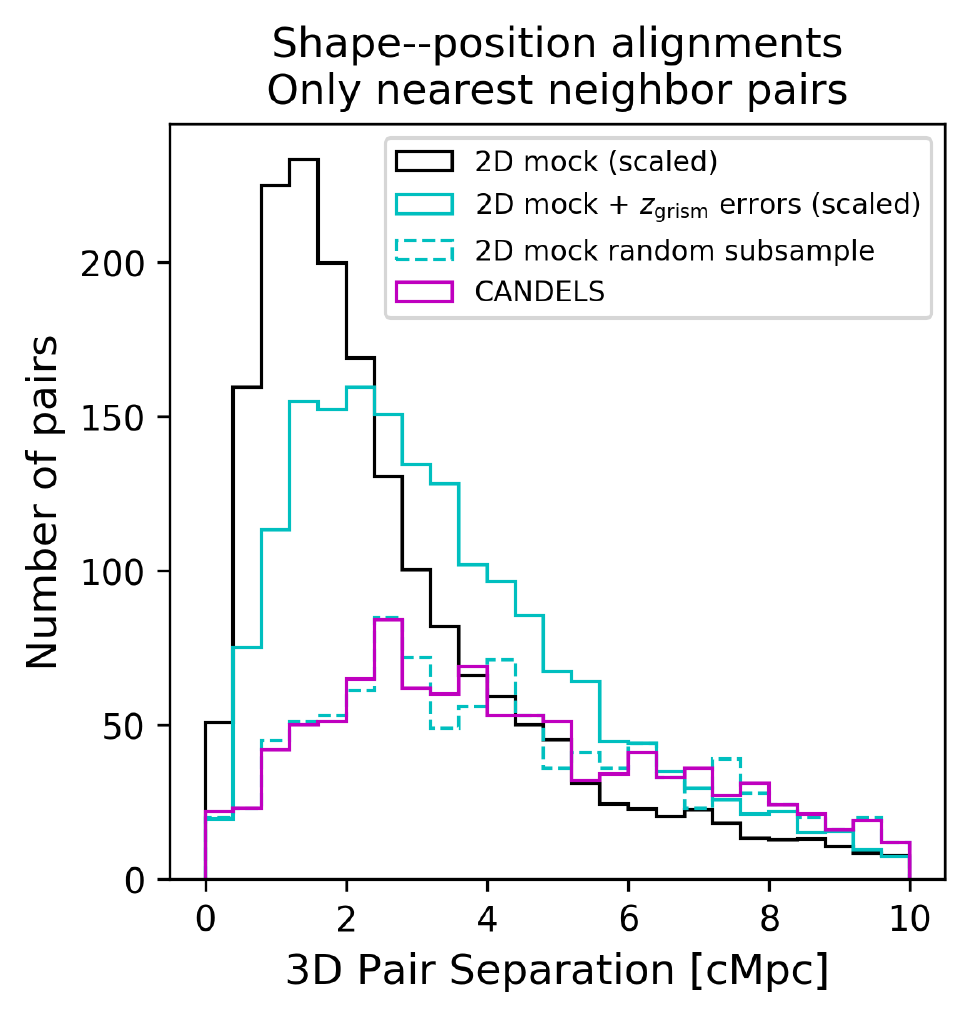}
\end{center}
\caption{Distributions of 3D comoving pair separations for nearest-neighbor shape--position alignment pairs. The overall 2D mock is shown in black (scaled down by a factor of five for clarity), and the same 2D mock but now with $z_{\rm grism}$ errors is shown as the solid cyan histogram (also scaled down by a factor of five for clarity). One random subsample of the $z_{\rm grism}$-error 2D mock is shown as the dashed cyan histogram; this subsample is designed to roughly match the observed number of pairs as a function of pair separation (magenta).}
\label{fig:sepns}
\end{figure}

In \autoref{fig:sepns}, we show as an example the distribution of 3D comoving pair separations for nearest--neighbor shape--position alignment pairs in the 2D mock and the observations. The distributions for the original 2D mock (black) and the 2D mock with $z_{\rm grism}$ errors (solid cyan) have both been scaled down by a factor of five for clarity. The $z_{\rm grism}$ errors clearly suppress the small separation pairs, and the peak in the distribution has shifted from $\sim1$ cMpc to $\sim2$ cMpc. The observed distribution is shown in magenta, and one random realization from our subsampling procedure is shown as the dashed cyan histogram; this random subsample closely matches the observed number of pairs as a function of pair separation. The fraction of such random subsamples that predict a signal will enable us to more fairly interpret the observational results. We note that an identical procedure was applied for the shape--shape alignment pairs and for the samples involving all neighbors within 10 cMpc. 

Given the disagreement between the mock and observed distributions of pair separations, it is useful to also compare angular two-point correlation functions using the full photometric CANDELS sample. We show in Appendix \ref{sec:appcorrfunc} that the correlation functions agree relatively well, although cosmic variance has a significant impact for ``pencil beam" surveys such as CANDELS.

\section{Results}\label{sec:results}
Here we report the results of our alignments analysis. We start with shape--position alignments followed by shape--shape alignments. For both types of alignments, we first present the expected signal from the mock simulations (starting with 3D real space, then degrading to 3D redshift space and 2D projected redshift space) followed by results from the CANDELS observations.

\subsection{Shape--position alignments}
First we discuss results for shape--position alignments.

\subsubsection{3D real space mocks}
In the top-most row of \autoref{fig:shapeposition}, we show the predicted signal from the mock simulations in 3D real space for shape--position alignments. The marginalized distribution of 3D alignment angles $|\cos\theta|$ is clearly non-uniform, with an excess toward $|\cos\theta|\sim1$. This effect is stronger when only the direction to the position of the nearest neighbor is considered, but the signal still persists when all neighbors within 10 cMpc are included. Indeed, the mean alignment angle as a function of 3D pair separation shows that the alignment is strongest for small pair separations on average. The $p$-value from a two-sample Kolmogorov-Smirnov (KS) test comparing each 3D mock $\cos\theta$ distribution to a random uniform distribution with the same number of data points is $\ll0.01$ for both nearest neighbor pairs and when considering all neighbors within 10 cMpc. This statistically supports our interpretation that the 3D real space mock exhibits a strong net alignment signal compared to the null hypothesis of no net excess of alignments.

\subsubsection{3D redshift space mocks}
In the second row of \autoref{fig:shapeposition}, we show an analogous plot for mock shape--position alignments but now after degrading from 3D real space to 3D redshift space (i.e., including the effects of RSDs). The distributions of 3D alignment angles $|\cos\theta|$ are still clearly non-uniform with an excess toward $|\cos\theta|\sim1$, indicative of net alignments. However, the RSDs have suppressed the peak at $|\cos\theta|\sim1$ and the strong dependence of $|\cos\theta|$ on pair separation is gone. The $p$-values for KS tests comparing the mock $\cos\theta$ distributions to a random uniform distribution for shape--position alignments are $\ll0.01$ for both nearest neighbors and all neighbors within 10 cMpc.

\subsubsection{2D projected redshift space mocks}
We now consider mock results in 2D projected redshift space for shape--position alignments (third row in \autoref{fig:shapeposition}). Since we are no longer working in 3D, we consider the 2D projected angle $\psi$ rather than the 3D angle $|\cos\theta|$. Statistically strong signals are predicted for shape--position alignments to nearest neighbors and to all neighbors within 10 cMpc. This is supported by KS tests comparing each 2D mock distribution to a random uniform distribution: the $p$-values are $\ll0.01$ for shape--position alignments to nearest neighbors and all neighbors within 10 cMpc.

The fourth row shows results for one random realization of our ``spectroscopically-degraded" 2D mock, where we include $z_{\rm grism}$ errors and randomly subsample the full 2D mock to match the observed number of pairs as a function of pair separation (see \autoref{sec:incompleteness}). It is immediately obvious that the alignment angle distribution for this particular realization is more noisy due to the smaller number of pairs, especially at small separations. Despite the increased noise in this particular realization, the signal is still statistically significant both when using nearest neighbor pairs and when considering all neighbors within 10 cMpc ($p\ll0.01$ in both cases). Of the 1000 random realizations, only 538 have $p<0.01$ (i.e., a statistically significant signal) when considering the shape--position signal for nearest neighbors, and $834/1000$ realizations have $p<0.01$ when using all neighbors (in comparison to a random uniform distribution).

\subsubsection{CANDELS observations}\label{sec:resultsobs}
\begin{figure*}
\begin{center}
\includegraphics[width=0.5\hsize]{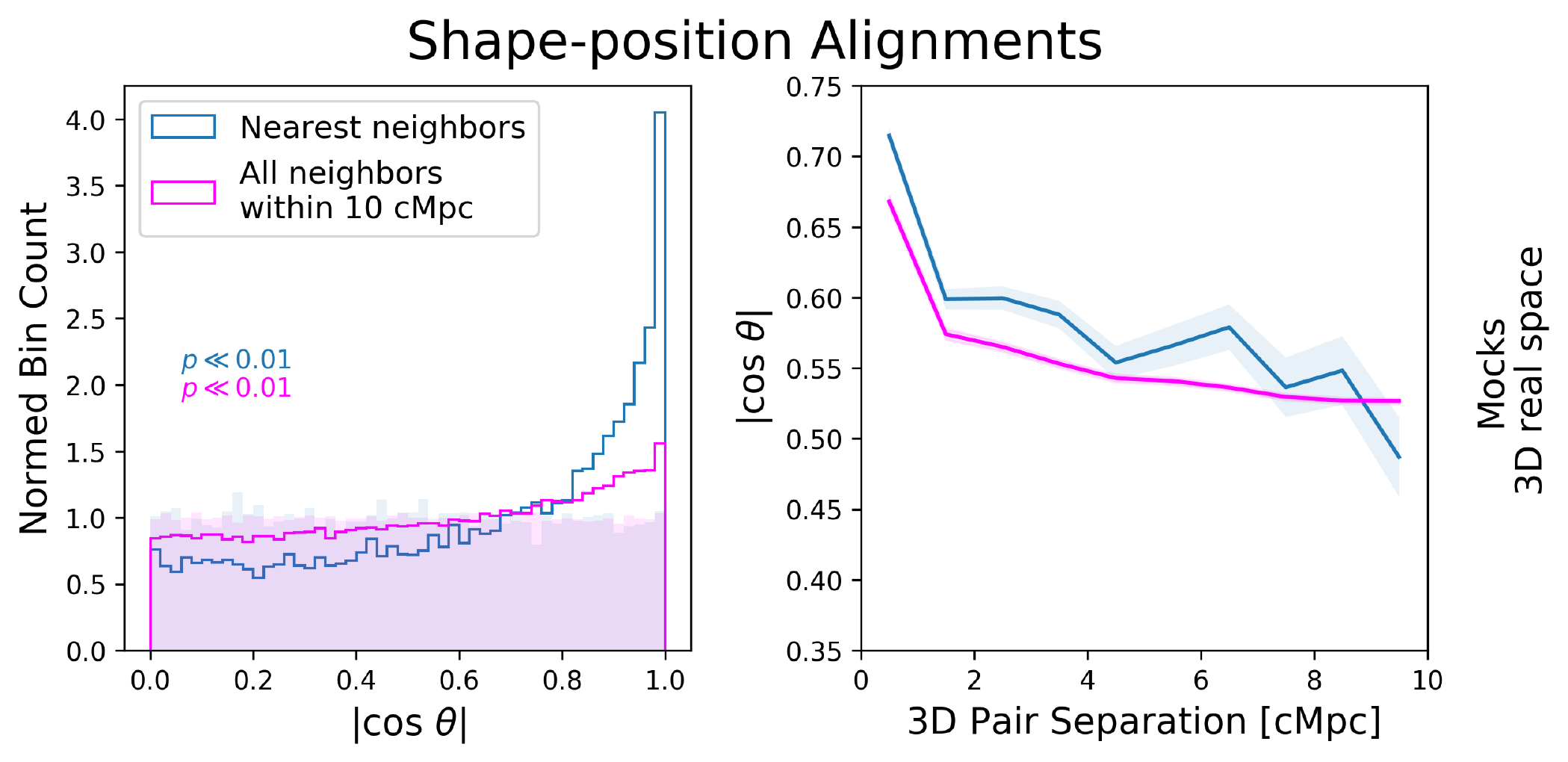}
\includegraphics[width=0.5\hsize]{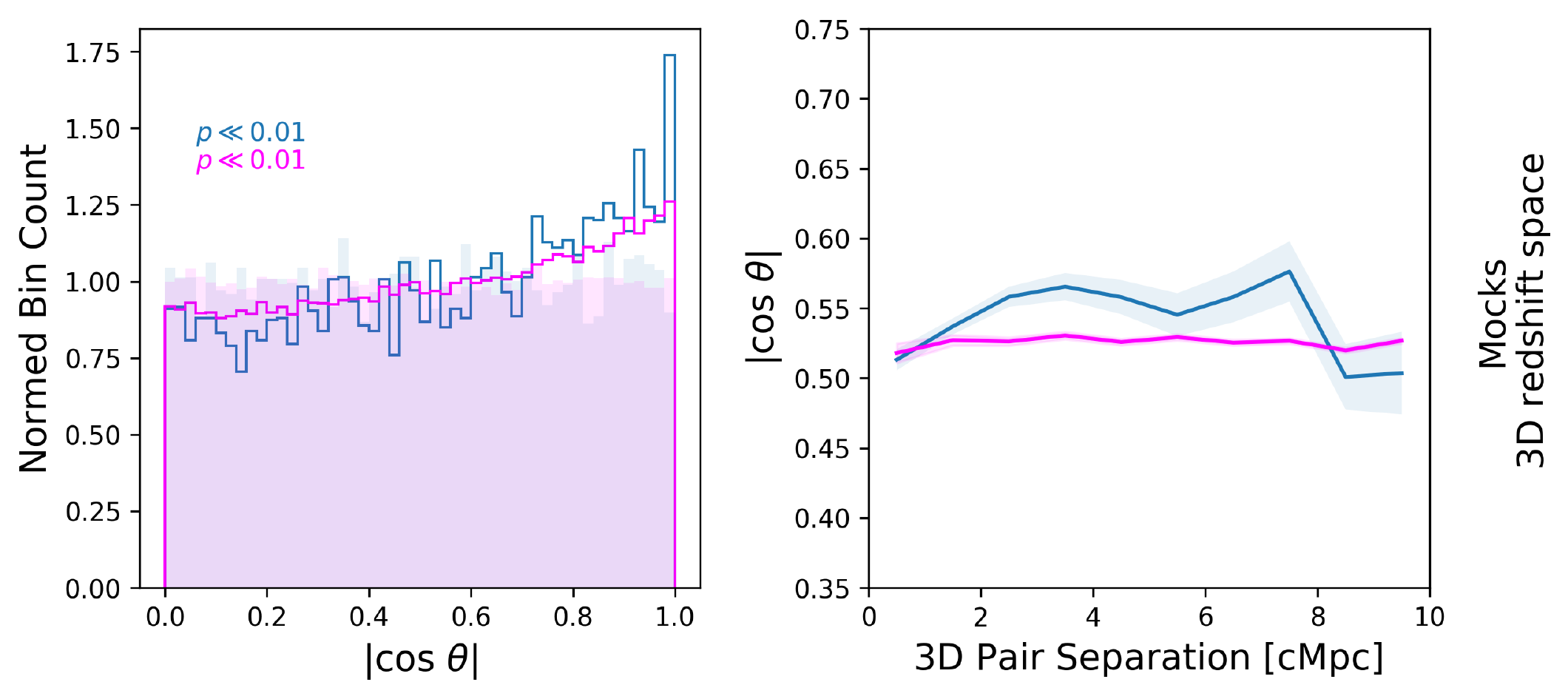}
\includegraphics[width=0.5\hsize]{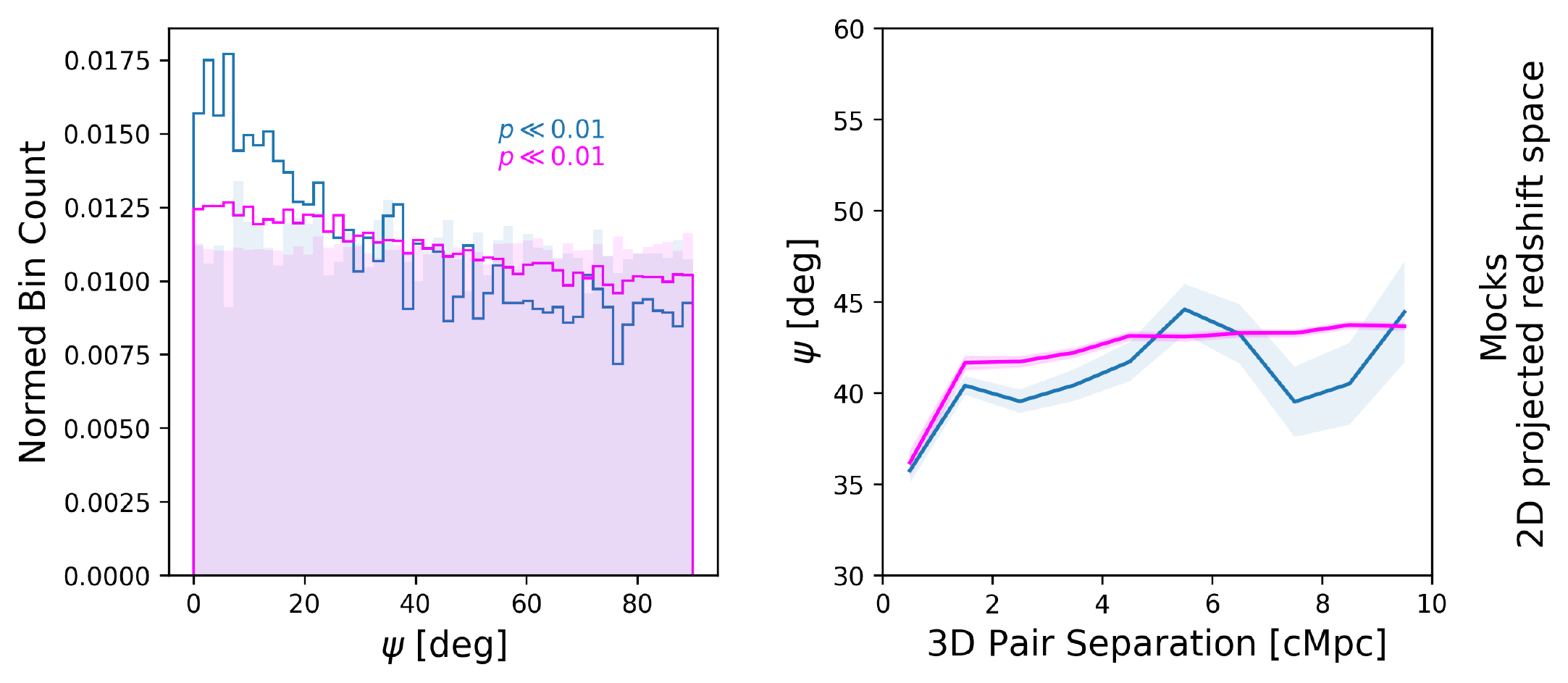}
\includegraphics[width=0.5\hsize]{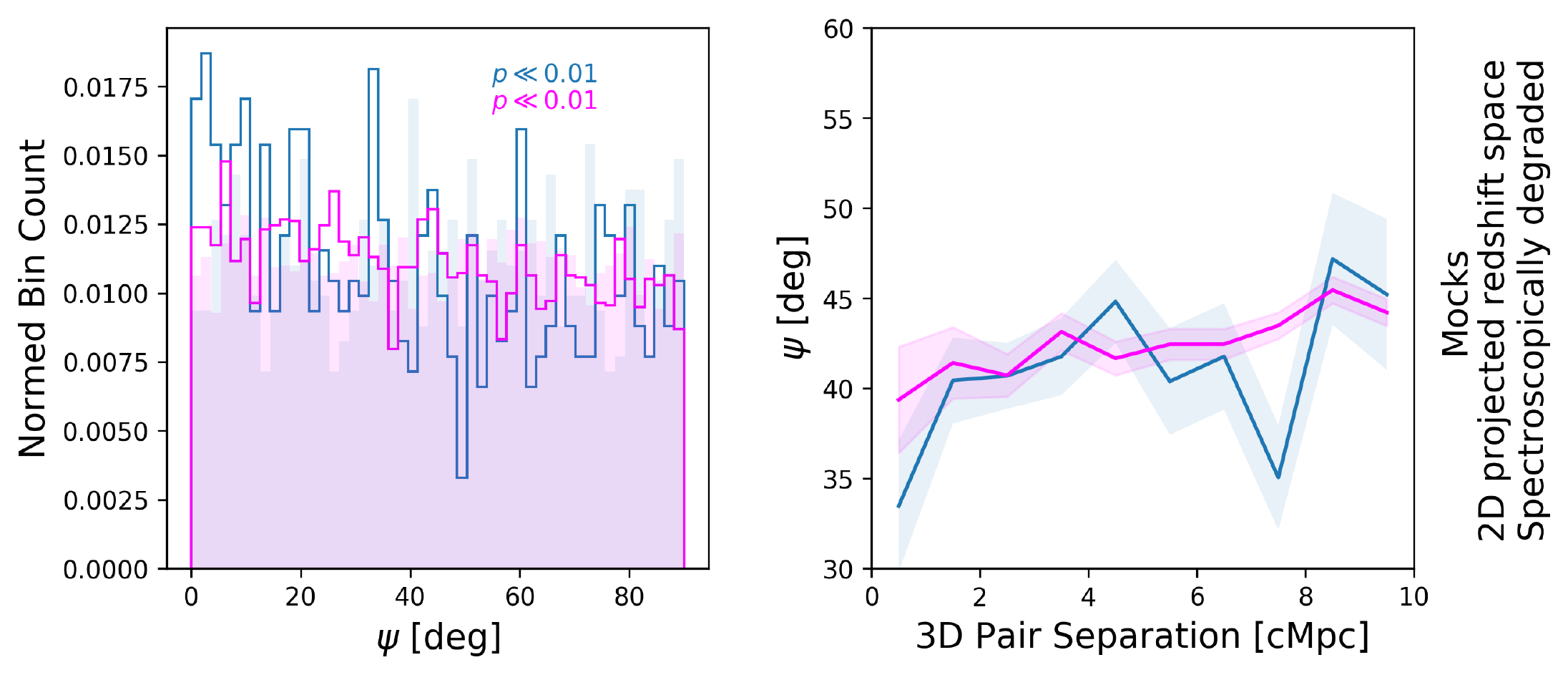}
\includegraphics[width=0.5\hsize]{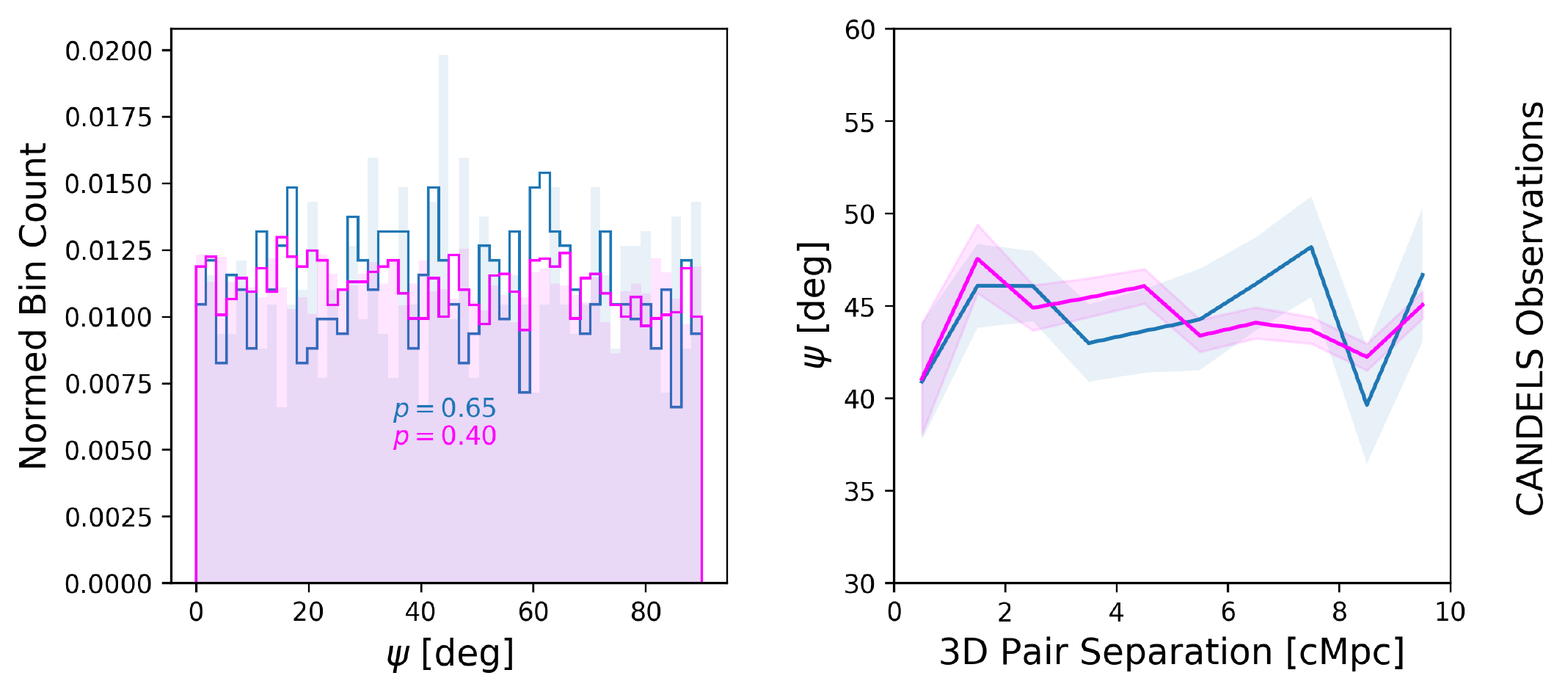}
\end{center}
\caption{Results for shape--position alignments. From top to bottom: 3D real space mocks, 3D redshift space mocks, 2D projected redshift space mocks, 2D projected redshift space mocks with spectroscopic incompleteness (see \autoref{sec:incompleteness}), and CANDELS observations. The left column shows the normalized alignment angle distributions for nearest neighbor pairs (blue), and all neighbors within 10 cMpc (magenta). The shaded histograms are random uniform distributions with the same number of data points as their respective (same color) comparison distributions. The right column shows how the mean alignment angle changes as a function of comoving pair separation, with the shading reflecting the standard error on the mean. In 3D, we plot the result of the dot product of two 3D vectors ($|\cos\theta|$), whereas in 2D we plot the difference between two projected angles ($\psi$). For the left panels, we write the $p$-values from KS tests comparing each distribution to the uniform distribution.}
\label{fig:shapeposition}
\end{figure*}

In the bottom row of \autoref{fig:shapeposition}, we show the distribution of 2D projected alignment angles $\psi$ for shape--position alignments in the CANDELS observations. It is evident that the marginalized distribution of $\psi$ is uniform regardless of whether we look only at nearest neighbors or include all neighbors within 10 cMpc. There is also clearly no trend in $\psi$ as a function of pair separation (e.g., smaller pair separation does not imply systematically lower $\psi$). Running a KS test comparing the observed distribution of shape--position alignment angles to a random uniform distribution yields $p$-values of 0.65 for nearest neighbors and 0.40 when considering all neighbors within 10 cMpc. These $p$-values are sufficiently large that we cannot rule out the null hypothesis that both sets of distributions are drawn from the same underlying parent population, i.e., there is insufficient evidence for a net excess of aligned pairs in the observations.

\subsection{Shape--shape alignments}
Now we discuss results for shape--shape alignments.

\subsubsection{3D real space mocks}
In the top row of \autoref{fig:shapeshape}, we show the predicted signal for shape--shape alignments in the 3D real space mocks. Here, both halos in a pair are assumed to host a prolate galaxy based on their redshift and assigned stellar mass. The signal is much weaker for these shape--shape alignments than for shape--position alignments, but a clear net excess of aligned pairs of $\cos\theta\sim1$ is visible when considering nearest neighbor pairs (indeed $p\ll0.01$ when comparing to a random uniform distribution with a KS test). While the signal appears weaker when considering all neighbors within 10 cMpc, it is still statistically significant in comparison to a random uniform distribution ($p\ll0.01$). Interestingly, $|\cos\theta|$ does appear to depend on pair separation, with smaller pair separations exhibiting slightly higher $|\cos\theta|$. 

\subsubsection{3D redshift space mocks}
In the second row of \autoref{fig:shapeshape}, we show the predicted signal for shape--shape alignments after degrading the mocks from 3D real space to 3D redshift space. Again, the signal seen in 3D real space is still present for nearest neighbors but not when considering all neighbors within 10 cMpc. Interestingly, while the shape--position alignments signal was degraded significantly in 3D redshift space compared to 3D real space, the shape--shape alignments signal appears only marginally weaker in 3D redshift space (although it is weaker to begin with in 3D real space). This is expected because RSDs will affect the positions of galaxies but not necessarily their orientations, hence the degradation from RSDs will be stronger for shape--position alignments than shape--shape alignments. The $p$-values for KS tests comparing the mock $\cos\theta$ distributions to a random uniform distribution for shape--shape alignments are $\ll0.01$ for both nearest neighbor pairs and when considering all neighbors within 10 cMpc.

\subsubsection{2D projected redshift space mocks}
In the third row of \autoref{fig:shapeshape}, we show the predicted signal for shape--shape alignments in the 2D projected redshift space mocks. A statistically significant net excess of shape--shape alignments is present in the 2D mock; a KS test comparing this mock distribution to a random uniform distribution results in a $p$-value of $\ll0.01$. But no statistically significant signal is present when using all neighbors within 10 cMpc (an analogous KS test yields a $p$-value of 0.33). 

The fourth row shows results for one random realization of our ``spectroscopically-degraded" 2D mock. Just as for the shape--position alignments, the obvious signal in the full 2D mock is now much more noisy due to the smaller number of pairs and $z_{\rm grism}$ errors. Indeed, for this particular realization, there is no statistically significant signal either when using only nearest neighbors ($p=0.14$) or all neighbors ($p=0.53$). Of the 1000 realizations, only 22 have $p<0.01$ when considering nearest neighbors, and only 15 realizations have $p<0.01$ when considering all neighbors (in comparison to a random uniform distribution).

\subsubsection{CANDELS observations}\label{sec:resultsobs}
\begin{figure*} 
\begin{center}
\includegraphics[width=0.5\hsize]{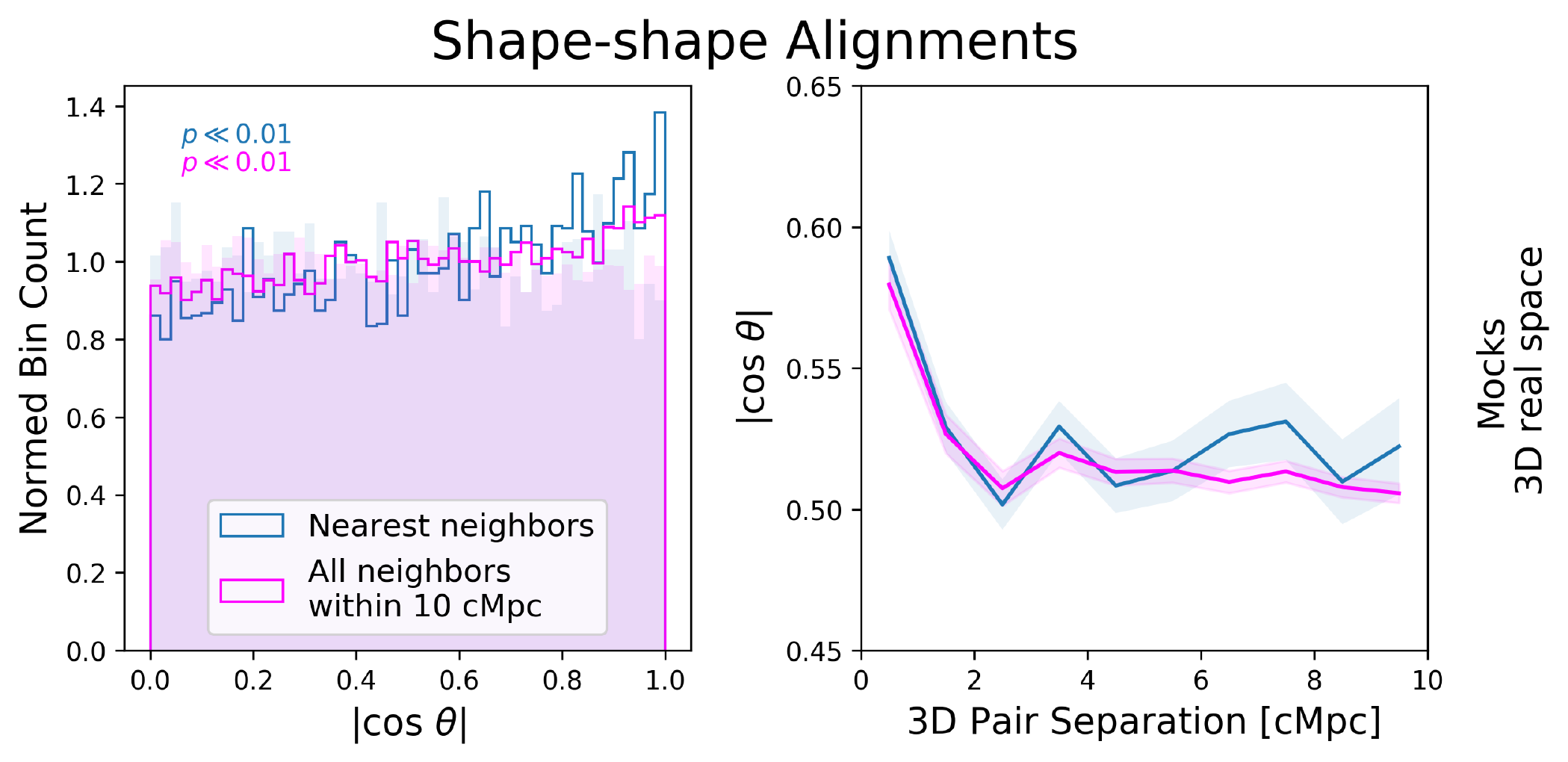}
\includegraphics[width=0.5\hsize]{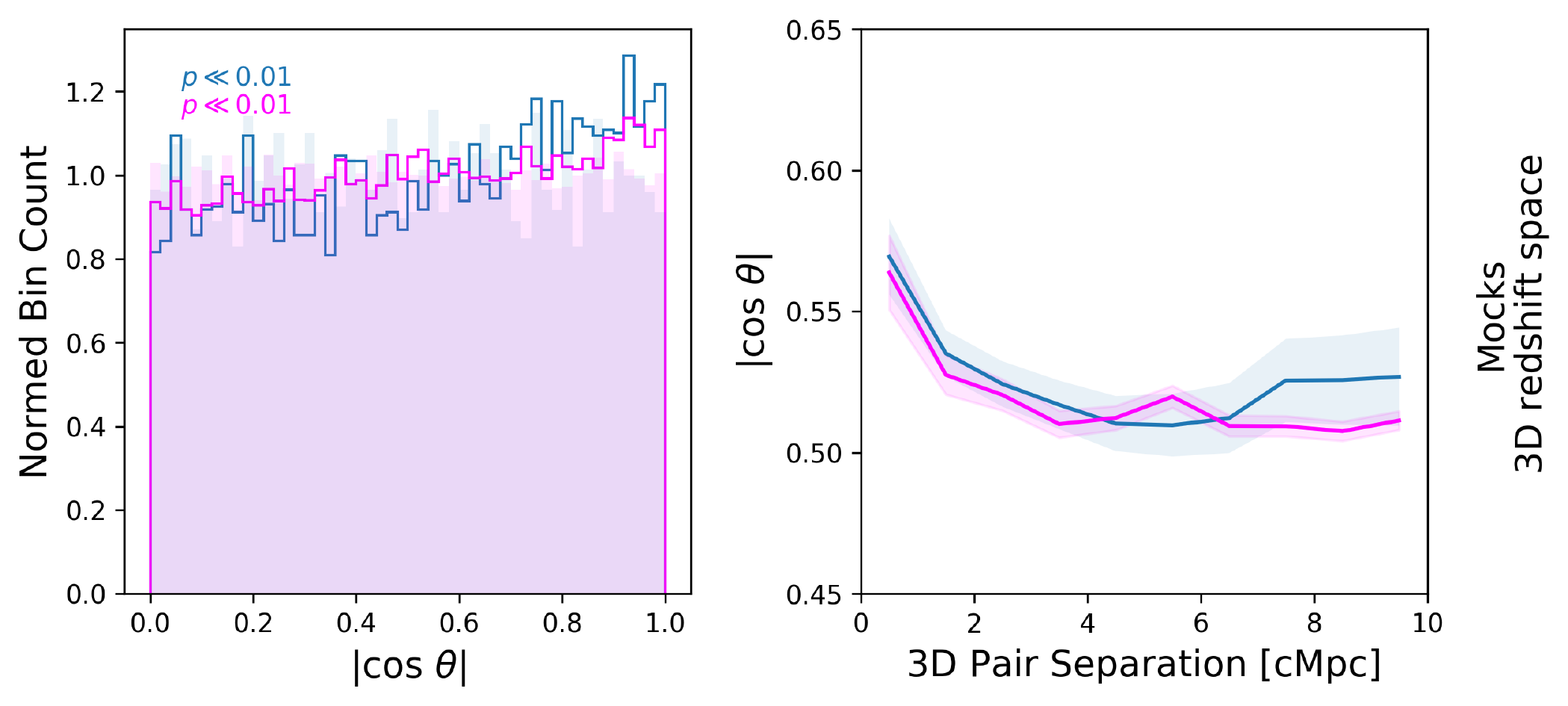}
\includegraphics[width=0.5\hsize]{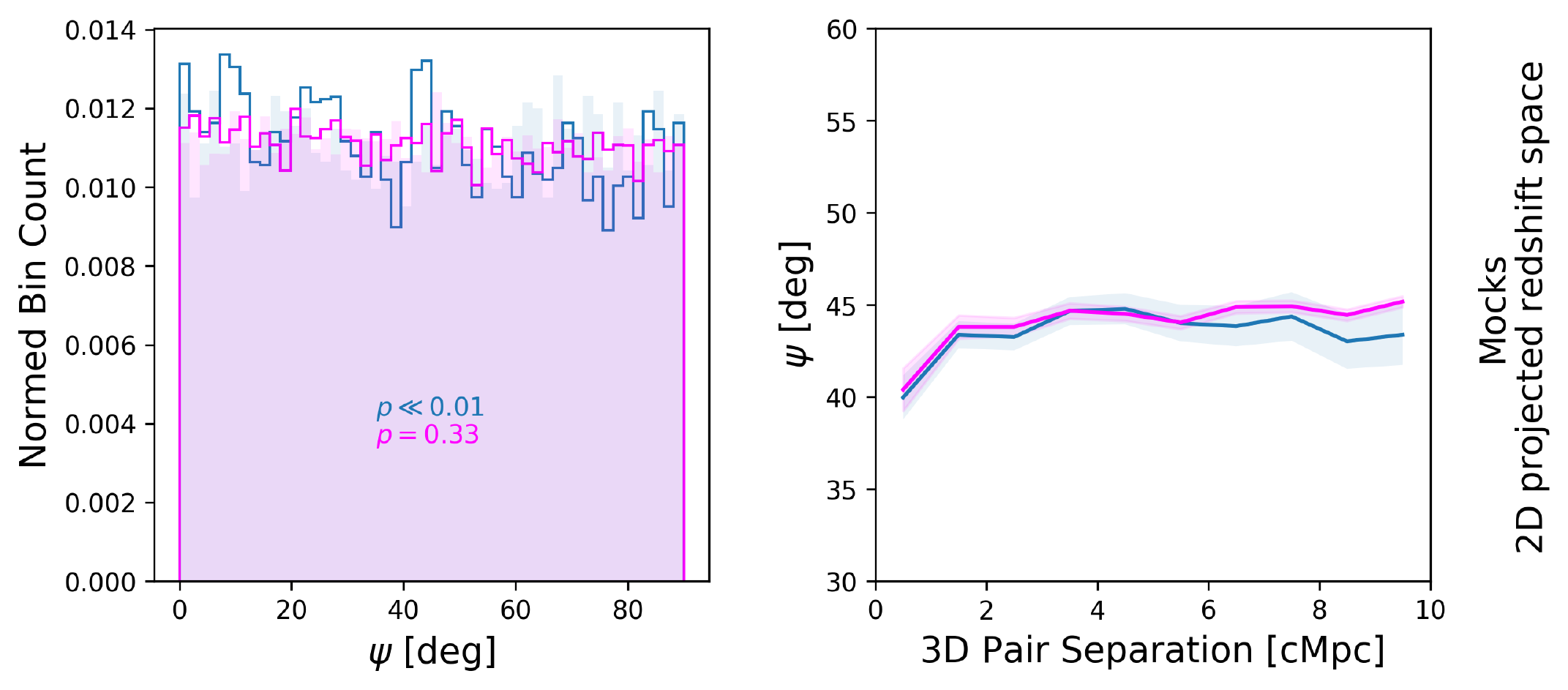}
\includegraphics[width=0.5\hsize]{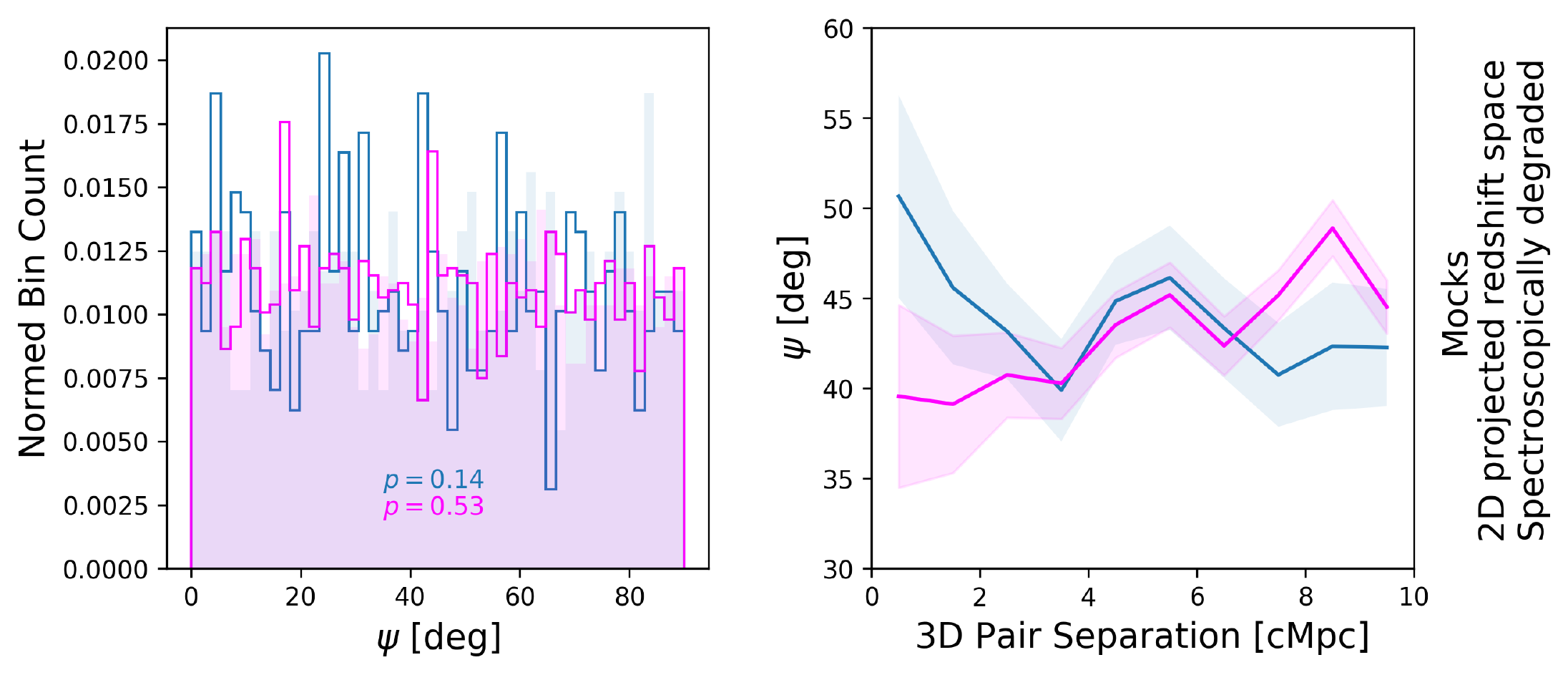}
\includegraphics[width=0.5\hsize]{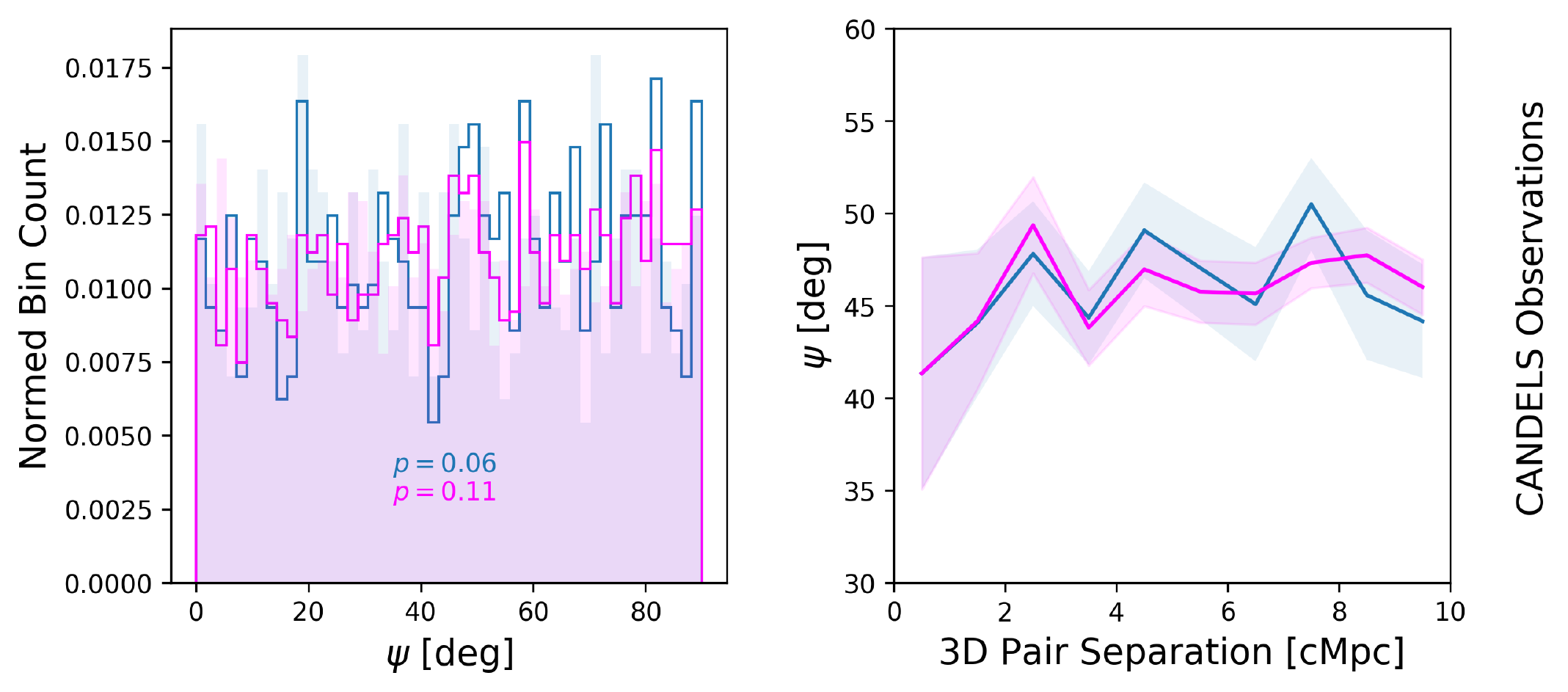}
\end{center}
\caption{Analogous to \autoref{fig:shapeposition} but now for shape--shape alignments.}
\label{fig:shapeshape}
\end{figure*}

In the bottom row of \autoref{fig:shapeshape}, we show the distribution of shape--shape alignment angles for the CANDELS observations. Again, there is no signal for shape--shape alignments in the CANDELS observations regardless of whether we consider only nearest neighbors or all neighbors within 10 cMpc. The $p$-values from a KS test comparing the observed distributions of shape--shape alignment angles to a random uniform distribution are 0.06 for nearest neighbors and 0.11 for all neighbors within 10 cMpc, both statistically insignificant. 

\section{Discussion}\label{sec:discussion}
Here we discuss our results and comment on the limitations of our analysis and prospects for future work.

\subsection{Interpreting the lack of an observed signal}
We showed that our analysis of the CANDELS observations does not reveal a signal for both shape--position and shape--shape alignments, regardless of whether we consider only nearest neighbors or all neighbors within 10 cMpc. In contrast, our mock lightcones predict a strong signal for both types of alignments in 3D real space, 3D redshift space, and 2D projected redshift space, particularly for nearest neighbors. However, when we account for grism redshift errors and spectroscopic incompleteness in the 2D mock, the predicted signal becomes much weaker. Considering only nearest neighbor pairs (where the predicted signal is strongest), only $\sim54\%$ of random realizations of the 2D mock show a statistically strong signal for shape--position alignments, and only $\sim2\%$ of random realizations show a strong signal for shape--shape alignments. We can also infer the probability that the observational data are consistent with the mock data by asking what fraction of the random mock realizations have $p$-values greater than the observed $p$-values. Of the 1000 realizations, only 10 have a larger $p$ value ($p>0.65$) for nearest neighbor shape--position alignments but 905 have $p>0.06$ for nearest neighbor shape--shape alignments. This indicates that the nearest neighbor shape--shape alignments in the mock data cannot be ruled out by the observational data, but that there is significant tension for nearest neighbor shape--position alignments.

\begin{figure} 
\begin{center}
\includegraphics[width=0.9\hsize]{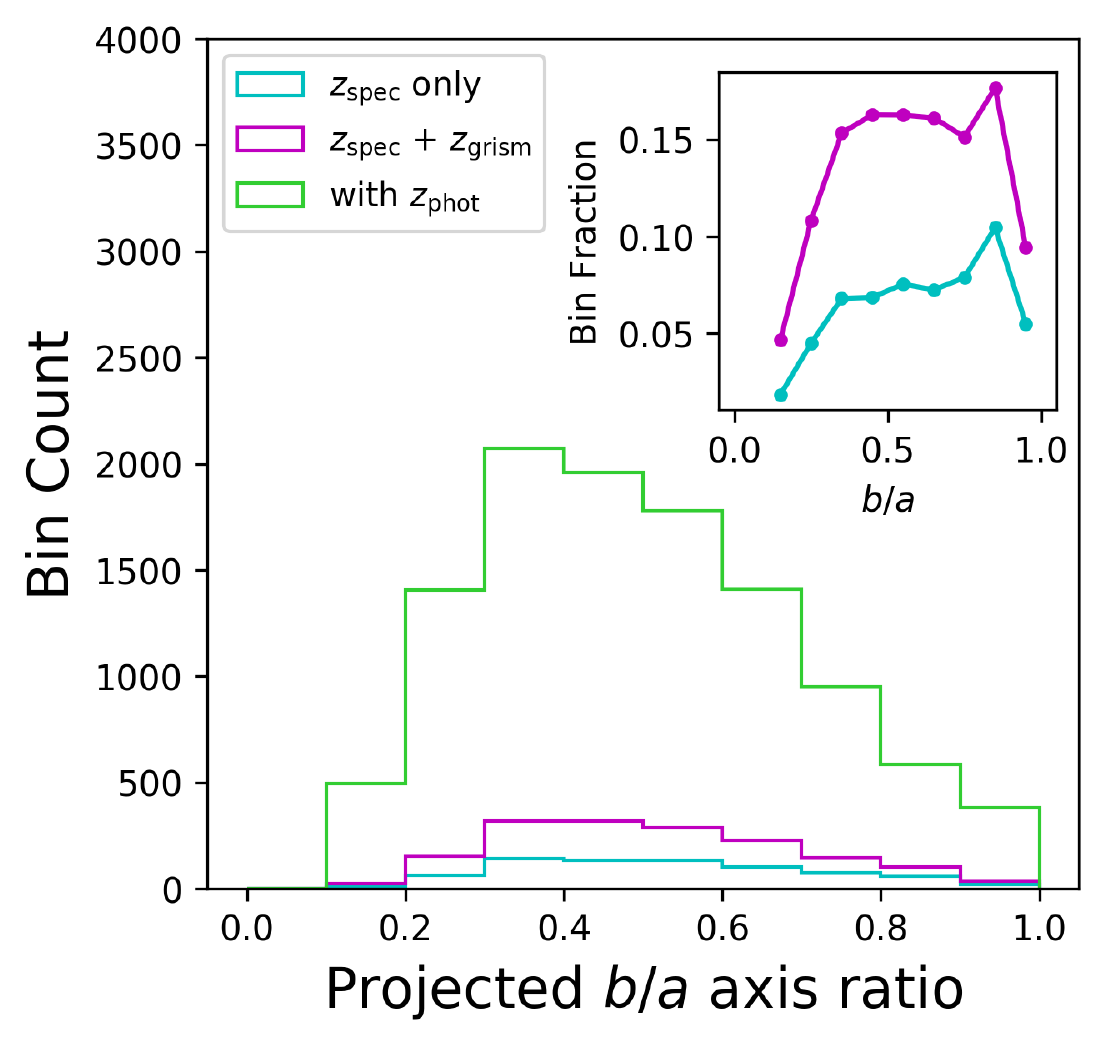}
\end{center}
\caption{Distributions of projected axis ratios for observed CANDELS galaxies with only high-quality spectra (cyan), with either high-quality or grism spectra (magenta), and with either spectroscopic or photometric redshifts (green). These distributions are for galaxies in our four CANDELS stellar mass-redshift bins, without any cut on galaxy shape or prolateness probability. The inset shows the fraction of galaxies that is spectroscopically-confirmed as a function of $b/a$. There appears to be a bias against obtaining spectroscopic redshifts for galaxies with low $b/a$, regardless of whether grism redshifts are included or not.}
\label{fig:zbestaxisratios}
\end{figure}

Our results suggest that increasing the sample size of the observations would help robustly constrain the expected signal (or lack thereof) since the shape--position and shape--shape signals for nearest neighbor pairs are both quite strong in the full 2D mock. The most efficient way to accomplish this would be to specifically target low-mass galaxies with $b/a\lesssim0.3$ at $z=1.0-2.5$ and their nearest neighbors in the photometric source catalogs. Indeed, in \autoref{fig:zbestaxisratios}, we plot the distribution of $b/a$ for galaxies in our four stellar mass--redshift bins with reliable GALFIT measurements, separately for the full photometric sample and the subset with spectroscopic or grism redshifts. The inset panel shows the fraction of galaxies that have spectroscopic or grism redshifts as a function of $b/a$. It is immediately obvious that there is an additional strong bias against obtaining spectroscopic (or grism) redshifts for low-mass elongated galaxies in CANDELS: the spectroscopic fraction in CANDELS is constant at the $\sim15\%$ level for intermediate $b/a$, but drops to $\sim5\%$ for $b/a<0.3$ (even when including grism redshifts). These are precisely the galaxies that are predicted to show strong internal galaxy--halo alignment and hence large-scale intrinsic alignments on scales of several cMpc \citep{tomassetti16,zhang18}. We have not attempted to mock this additional spectroscopic bias against low-mass galaxies with small $b/a$ since it is not understood how halo $b/a$ relates to galaxy $b/a$ as a function of halo mass and redshift. However, it is likely that if we had mocked this additional bias, then an even smaller fraction of our random realizations of the 2D mock would predict a statistically strong signal, and a larger fraction of realizations would have $p>0.79$ for nearest neighbor shape--position alignments, making the mock more consistent with the observational data. 

Another way to further degrade the expected mock signal is to allow for some scatter between prolate galaxy orientation and prolate halo orientation. The advantage of prolate galaxies as tracers of large scale structure via their intrinsic alignments is that they are nearly always aligned with their host halos on scales of $R_{\rm vir}$, modulo some small scatter according to \citet{tomassetti16}. Hence, prolate galaxy alignments should closely follow prolate halo alignments, which is not necessarily true for halos that host disks or spheroids. Indeed, one more way that the signal expected for prolate galaxies could be diluted or perhaps even countered is if disks are preferentially misaligned with their host halos \citep[due to their angular momentum vectors changing direction on short timescales during the early turbulent merger-driven phase of galaxy formation; e.g.,][]{martig14,ceverino14}. Although we have attempted to use only the most likely prolate galaxies (low $b/a$ and large size), it is possible that our expected signal in the observations is also affected by contamination from edge-on disks at some level. Finally, given the small sizes of the CANDELS fields, cosmic variance may have a significant impact on our results, although as we show in Appendix \ref{sec:appcorrfunc}, the redshift distributions of our mock and CANDELS datasets agree relatively well. It is likely that all of these effects are at play observationally, and should be considered in future analyses of hydrodynamical simulations and for making more realistic mock predictions.

\subsection{Limitations and future work}
Here we list several limitations of our analysis and discuss prospects for future work. First and foremost, our strong mock prediction is based on N-body dark matter-only simulations and hence our predictions reflect halo--halo alignments in the underlying $\Lambda$CDM cosmology. In the future, it will be useful to confront the predictions from our dark matter-only mocks with those from large-volume cosmological simulations with and without baryons (of sufficiently high resolution). That will provide insight into whether galaxies in the elongated phase are indeed almost always aligned with their dark matter halo on scales of $R_{\rm vir}$, whether elongated galaxy--galaxy alignments show roughly the same signal as the parent halo--halo alignments, and whether baryonic effects can significantly change the halo--halo alignments expected within the $\Lambda$CDM cosmology. It would also be useful to compare the proximity and orientation of prolate galaxies relative to nearby filaments in such hydrodynamical simulations. Another aspect to follow up on is whether the radius within which the dark matter halo shape is measured affects the intrinsic alignments signal; we have used halo shapes measured within $R_{\rm vir}$ as is common practice, but it is possible that the best-fitting halo ellipsoidal model would differ on smaller or larger scales \citep[say $R_{\rm 500c}$ or $2R_{\rm vir}$; e.g., see section 2.2 of][]{kiessling15}. Also related is whether baryonic effects in hydrodynamical simulations significantly change the halo shape (notably projected $b/a$), especially for the prolate phase; such results would be useful for semi-analytic models and subhalo abundance matching models that attempt to predict galaxy shape and orientation \citep[e.g., see][]{tenneti15,chisari15}.

On the observational side, it will be crucial to obtain spectroscopic redshifts for more galaxies with $b/a\lesssim0.3$ to more robustly constrain the expected alignment signal at small pair separations. In addition, cross-correlating our observed alignment angle measurements with other more traditional approaches to mapping the cosmic web may also be fruitful \citep[such as estimates of the density field and identifications of filamentary overdensities in redshift slices, which would enable probing ``shape--filament" alignments; e.g.,][]{darvish17,laigle18,lee18}. Incorporating environmental information to exclude galaxies in high-density regions may also be a useful way to mitigate the impact of RSDs on shape--position alignments and hence recover a signal. Furthermore, our characterization of the shapes of galaxies is based on the GALFIT structural catalogs of \citet{vanderwel12}. It would be useful to revisit the basic structural properties of these low-mass prolate galaxies with more sophisticated techniques since it is known that the intrinsic alignment signal is very sensitive to how the shapes are measured \citep[e.g.,][]{singh16}. It might be fruitful to try fitting ellipticity as a function of radius (in one or more bands) and using non-parametric approaches to determine the surface brightness profile shape rather than assuming single-component Sersic profiles. These more sophisticated measurements could then be folded into more formal approaches for measuring correlations between galaxy shapes as a function of pair separation, as is often done in weak gravitational lensing studies using non-linear alignment models \citep[e.g.,][]{hirataseljak04,mandelbaum11,joachimi11,johnston18}. Finally, it may also be possible to take advantage of the larger photometric sample by using different techniques that are robust against photometric redshift uncertainties (e.g., using on-sky angular pair separations in wide redshift slices rather than 3D comoving pair separations).

\section{Summary}\label{sec:summary}
In this paper, we used CANDELS observations and mock lightcones to measure intrinsic alignments for low-mass ($\log M_*/M_{\odot}=9-10$) elongated galaxies at high redshift ($z=1.0-2.5$). We constrained two types of alignments: (1) ``shape--position" alignments test whether the position angle of a galaxy points in the direction to its nearest neighbors, and (2) ``shape--shape" alignments test whether the position angles of two nearby low-mass prolate galaxies themselves are aligned in the same direction. We considered results separately for nearest neighbors only and for all neighbors within 10 cMpc. We restricted our observational analysis to only spectroscopically-confirmed galaxy pairs to minimize errors in 3D comoving pair separations. Our conclusions are as follows: 

\begin{enumerate}
\item Our mocks predict strong signals for both ``shape--position" and ``shape--shape" intrinsic alignments in 3D real space, and these statistically significant signals persist even after degrading to 3D redshift space and 2D projected redshift space. The signals are strongest when considering only nearest neighbor pairs rather than averaging over all neighbors within 10 cMpc.
\item We do not detect the signal predicted by the full 2D mock in the CANDELS observations for both shape--position and shape--shape alignments. 
\item When we ``spectroscopically degrade" the 2D mocks by accounting for grism redshift errors and generating 1000 random realizations where the number of pairs as a function of pair separation matches the observations, the alignment signals are severely degraded. Only $\sim54\%$ of realizations predict a statistically significant shape--position signal for nearest neighbors, and only $\sim2\%$ of realizations predict a statistically significant shape--shape signal for nearest neighbors. 
\item Hence, it is possible that spectroscopic incompleteness and biases are responsible for the lack of a signal in the observations. Of the many possible avenues for future work that we have highlighted, it may be most useful to obtain spectroscopic redshifts for more galaxies with $\log M_*/M_{\odot}=9-10$ and $b/a\lesssim0.3$ at $z=1.0-2.5$, and their nearest on-sky neighbors selected from the photometric source catalogs. 
\end{enumerate}

\section*{Acknowledgements}
We greatly thank the following people for helpful discussions about this project: Rachel Somerville, Doug Hellinger, Aldo Rodriguez-Puebla, Fangzhou Jiang, Nir Mandelkar, Christina Williams, Steve Finkelstein, David Spergel, Shy Genel, Ari Maller, Dick Bond, Kevin Bundy, Renbin Yan, Alexie Leauthaud, Bahram Mobasher, Marc Huertas-Company, Sandro Tacchella, Timothy Carleton, Kameswara Mantha, Harry Johnston and Neta Bahcall. We also thank the anonymous referee for helpful suggestions. We thank NASA Advanced Supercomputing for access to their Pleiades supercomputer where the Bolshoi--Planck simulations were run, and where we performed part of the analysis for this project. This material is based upon work supported by the National Science Foundation Graduate Research Fellowship Program under Grant No. 1339067 to VP, and by an HST grant to JP.

%%%%%%%%%%%%%%%%%%%%%%%%%%%%%%%%%%%%%%%%
\bibliographystyle{mnras}
\bibliography{references}
%%%%%%%%%%%%%%%%%%%%%%%%%%%%%%%%%%%%%%%%

\appendix

\section{Cosmic Variance}\label{sec:appcorrfunc}
Throughout this work, we have used a single mock lightcone to generate predictions for alignment angle distributions (for the sake of convenience). It is worthwhile to show that our results would be qualitatively similar if we used other mock lightcones. As mentioned in \autoref{sec:datamocks}, there exist eight lightcones for each of the five CANDELS fields, leading to forty different mock lightcone realizations with the UniverseMachine \citep{behroozi18}. Note that each mock lightcone represents an area on the sky larger than all five CANDELS fields combined, and hence we need to extract five smaller sized subfields as shown in \autoref{fig:subfields}. Analyzing all forty mock lightcones would be prohibitive, and hence here we simply compare our main mock lightcone results to three alternatives analyzed in the same way.\footnote{Specifically, for the three alternative mock light cones, we use survey\_GOODS-S\_z0.00-10.00\_x39.00\_y41.00\_1.dat, survey\_GOODS-S\_z0.00-10.00\_x39.00\_y41.00\_2.dat, and survey\_UDS\_z0.00-10.00\_x36.00\_y35.00\_4.dat.}

\begin{figure*} 
\begin{center}
\includegraphics[width=\hsize]{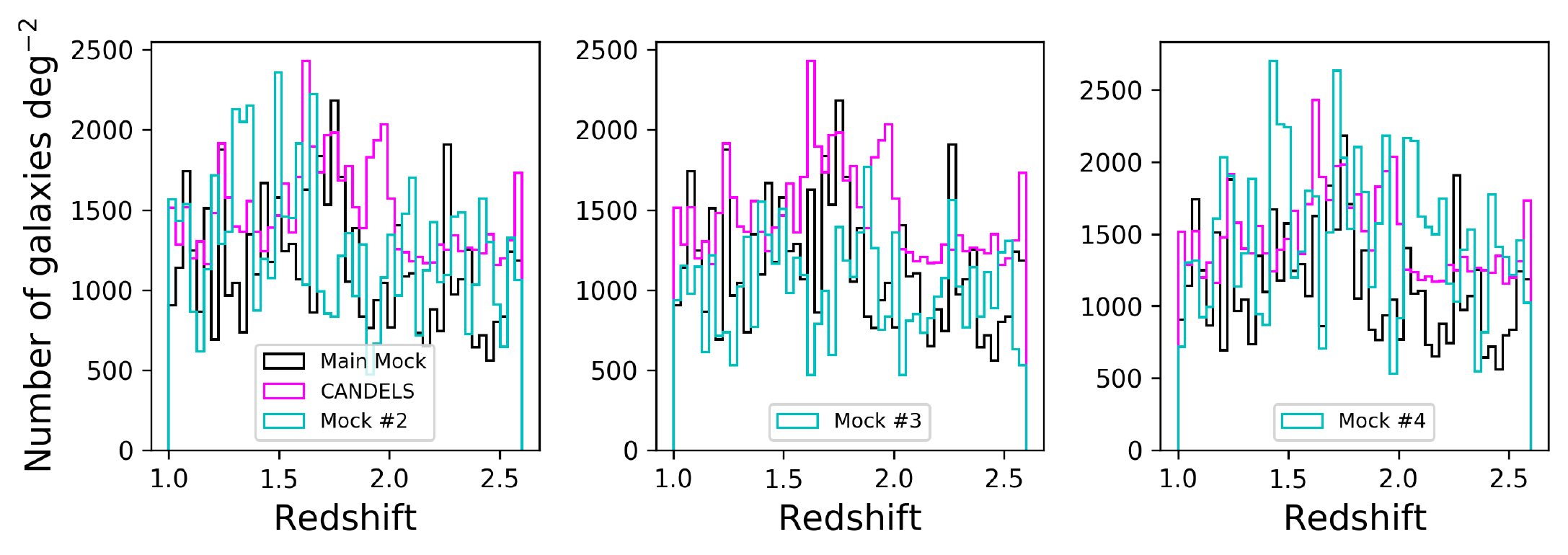}
\end{center}
\caption{The on-sky number density of galaxies with $\log M_*/M_{\odot}=9-10$ as a function of redshift for CANDELS (magenta), our main mock lightcone (black), and three alternative mock lightcones (cyan in the different panels). As expected, each mock has a different redshift distribution but the deviations are reasonable and broadly consistent with CANDELS.}
\label{fig:redshifts}
\end{figure*}

First, we show in \autoref{fig:redshifts} the on-sky number density of galaxies as a function of redshift for CANDELS observations and the different mocks. We restrict this comparison to galaxies with $\log M_*/M_{\odot}=9-10$. Our main mock agrees relatively well with CANDELS, which is to be expected on average since the mocks are generated within the framework of subhalo abundance matching. However, we specifically point out that the main mock shows a deficit of galaxies in the redshift range $z=1.8-2.2$ compared to CANDELS, which probably reflects a larger abundance of voids along the line-of-sight in this particular mock. The redshift distributions of the three alternative mocks show reasonable deviations from each other, from the main mock, and from CANDELS. For example, the alternative mock in the rightmost panel of \autoref{fig:redshifts} shows an excess spike in the number of galaxies relative to CANDELS at $z=1.8-2.2$ unlike in our main mock.

We showed in \autoref{sec:incompleteness} that the observed spectroscopic pair sample has a deficit of smaller separation pairs compared to the 2D mock. While this is likely due to the complicated and unknown spectroscopic follow-up selection function (and hence spectroscopic incompleteness) for CANDELS, it is worthwhile to ask whether the angular two-point correlation functions agree between the mocks and the full underlying CANDELS sample (i.e., including galaxies with photometric redshifts). This is especially interesting because the subhalo abundance matching framework of the UniverseMachine was calibrated to match observed correlation functions only at $z\lesssim0.7$ \citep{behroozi18}, and hence there is no guarantee that the mock correlation functions will agree at higher redshift \citep[but see, e.g.,][]{kravtsov04}. Hence, here we compute the angular two-point correlation functions for all five CANDELS fields and the mocks. 

It is customary to compute correlation functions in stellar mass bins and redshift bins. Here we consider only one stellar mass bin ($\log M_*/M_{\odot}=9-10$) and four redshift bins that are relevant for our study ($z=1.0-1.4, 1.4-1.8, 1.8-2.2, 2.2-2.6$). We do not try to account for the complicated spectroscopic selection function and its effect on biasing the correlation function; instead we use the combined photometric and spectroscopic ``$z_{\rm best}$" sample. Specifically, we use spectroscopic redshifts where available and photometric redshifts otherwise. The width of the redshift bins ($\Delta z=0.4$) ensures that a significant number of galaxies cannot scatter into or out of our redshift bins due to photometric redshift uncertainties. Our mock correlation functions are computed independently in five subfields whose rectangular areas are equivalent to the CANDELS fields (see \autoref{fig:subfields}). Note that the real CANDELS fields have jagged outer edges due to the way the HST observations were tiled. We do not attempt to mock that effect here. Future papers that focus exclusively on CANDELS correlation functions may be able to deal with this problem in a simple way by extracting slightly smaller sized rectangles from the overall observed CANDELS fields, thereby excluding the jagged outer edges, or by using the more sophisticated approach of \citet{landyszalay93} which gives excess pair counts above the expected pair counts for randomly distributed sources (and hence is more robust against biases due to survey geometry). 

\begin{figure*} 
\begin{center}
\includegraphics[width=\hsize]{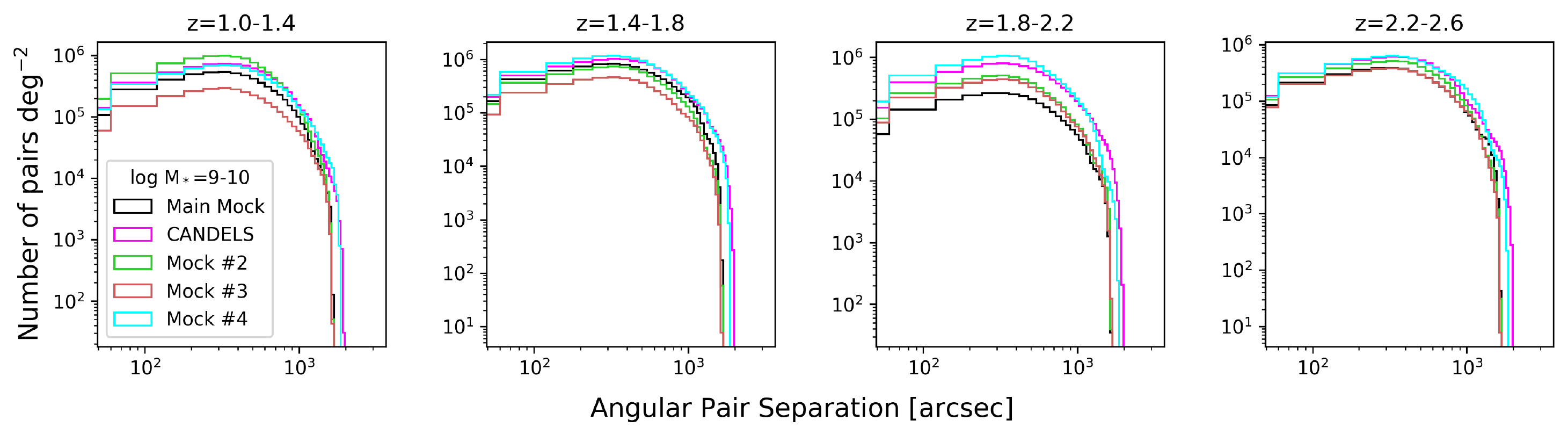}
\end{center}
\caption{Angular two-point correlation functions for the CANDELS observations (magenta), our fiducial mock (black), and three alternative mocks. From left to right are the correlation functions in each of our four redshift bins. These correlation functions are only for galaxies with $\log M_* / M_{\odot}=9-10$. The disagreements between the various mocks and with CANDELS are likely due to cosmic variance as suggested by \autoref{fig:redshifts}.}
\label{fig:corrfunc}
\end{figure*}

In \autoref{fig:corrfunc}, we show the angular two-point correlation functions for CANDELS and the different mocks in each of our redshift slices for galaxies with $\log M_*/M_{\odot}=9-10$. The observed and fiducial mock correlation functions agree relatively well, although the main mock has a lower amplitude for the $z=1.8-2.2$ redshift bin, coinciding with its lower number of galaxies in the same redshift range seen in \autoref{fig:redshifts}. Similarly, the correlation functions of the alternative mocks qualitatively also look reasonable, with differences in amplitudes coinciding with excesses or deficits in their respective redshift distributions (relative to each other) as expected. In other words, the differences in the correlation function amplitudes qualitatively appear to arise from cosmic variance. Further quantifying cosmic variance is beyond the scope of this paper but this exercise shows that it can have a significant effect for ``pencil beam" surveys such as CANDELS. That the overall shapes of the CANDELS and mock angular two-point correlation functions agree relatively well, especially at small separations, is encouraging, however.

\begin{figure*} 
\begin{center}
\includegraphics[width=\hsize]{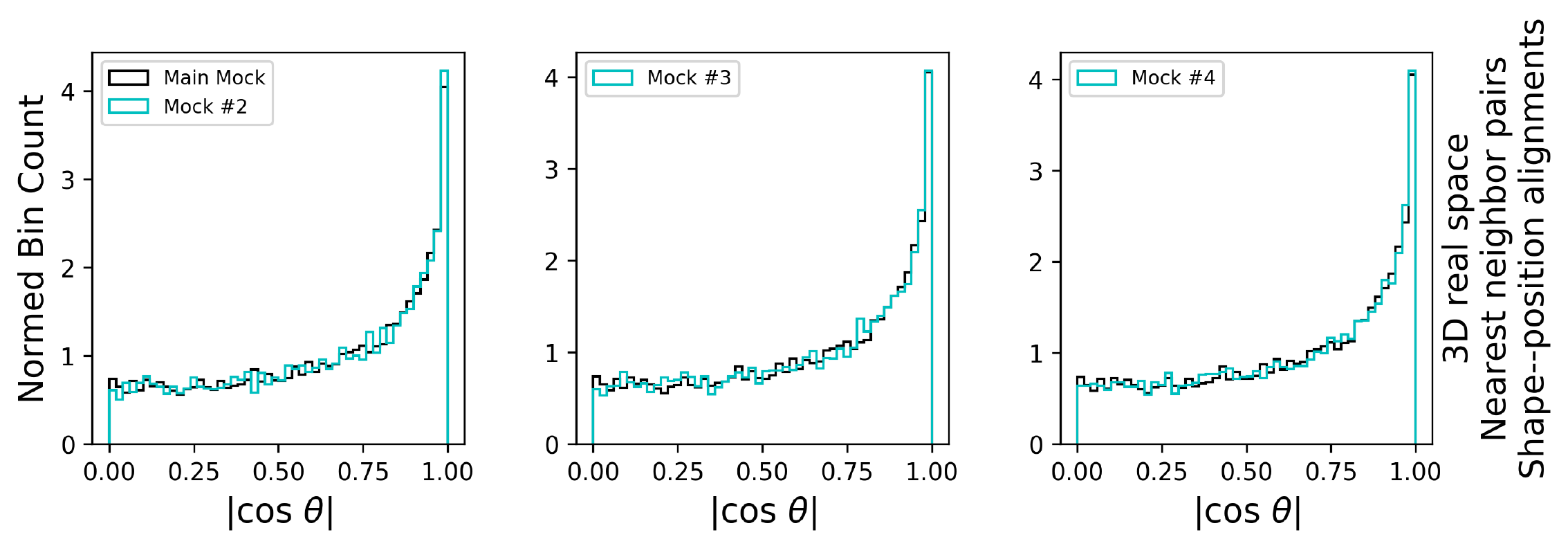}
\includegraphics[width=\hsize]{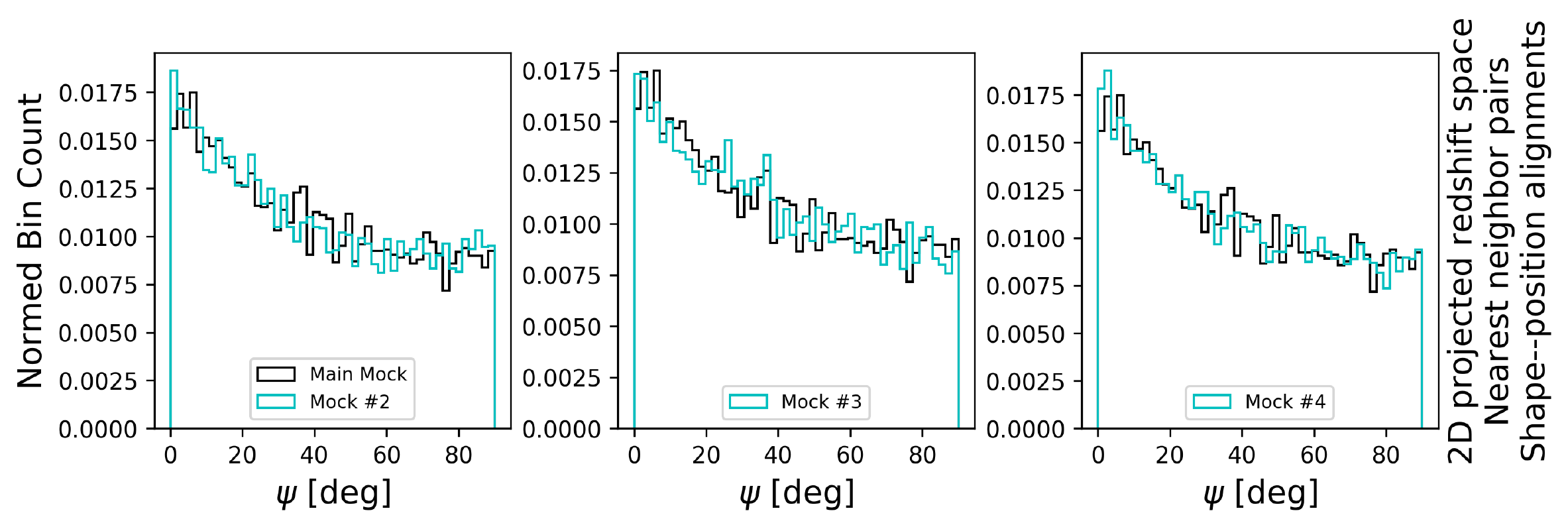}
\end{center}
\caption{Shape--position alignment angle distributions in 3D real space (top row) and 2D projected redshift space (bottom row) for nearest neighbor pairs in the three alternative mocks. The results are qualitatively similar when comparing the distribution of each alternative mock to that of our main mock.}
\label{fig:appshapeposition}
\end{figure*}

\begin{figure*} 
\begin{center}
\includegraphics[width=\hsize]{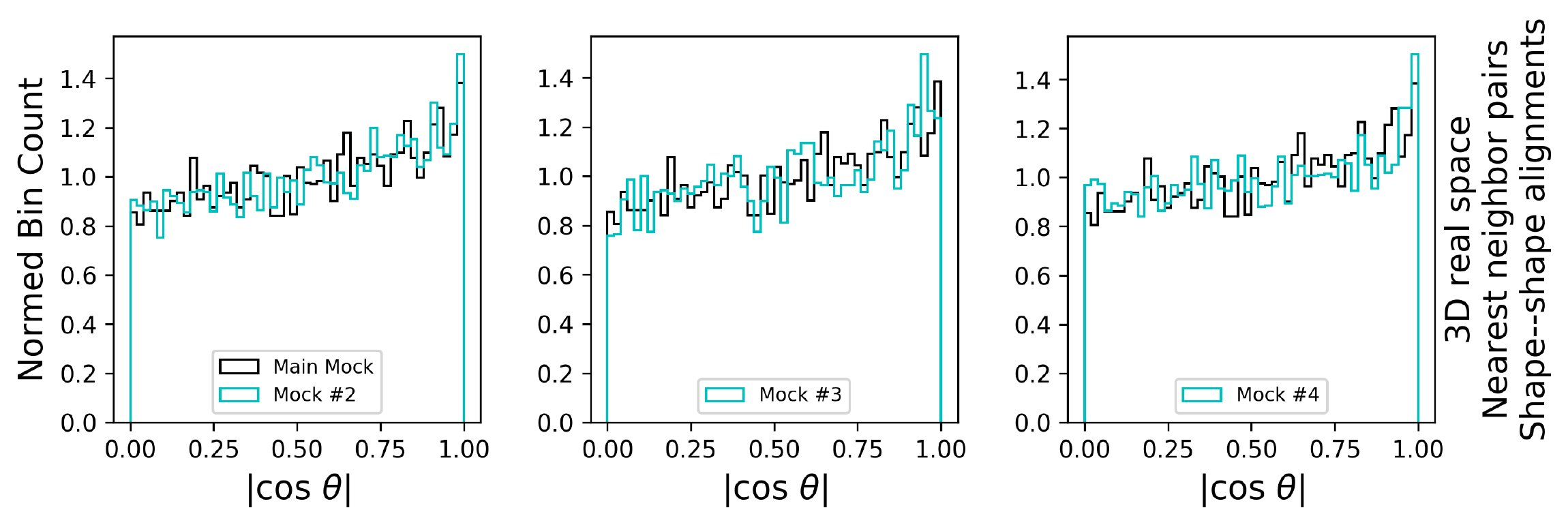}
\includegraphics[width=\hsize]{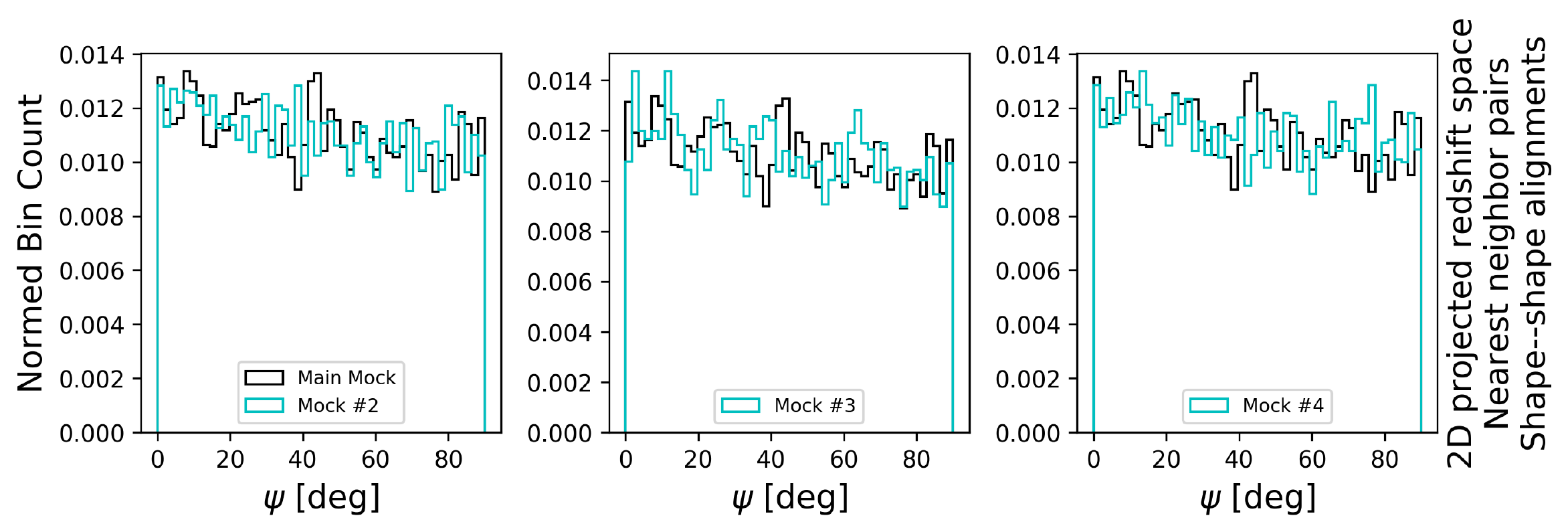}
\end{center}
\caption{Analogous to \autoref{fig:appshapeposition} but now for shape--shape alignments.}
\label{fig:appshapeshape}
\end{figure*}

Finally, we can compare the distributions of alignment angles across the different mocks. For the sake of convenience, we restrict our comparison to nearest neighbor pairs (for which the signals in the main mock are strongest) and only show results for 3D real space and 2D projected redshift space. In \autoref{fig:appshapeposition}, we show results for shape--position alignments. It is clear that despite the alternative mocks having different redshift distributions and pointing in different directions within the simulation box frame, their alignment angle distributions agree qualitatively well with those of the main mock (in both 3D real space and 2D projected redshift space). \autoref{fig:appshapeshape} shows that this is also true for shape--shape alignments for nearest neighbor pairs in both 3D real space and 2D projected redshift space. Hence, we do not expect that our conclusions drawn from the main mock would change had we used one or multiple alternative mocks for this work.

\end{document}